\documentclass[twocolumn]{aastex63}

\usepackage{graphicx}
\usepackage[T1]{fontenc}
\usepackage{aecompl}
\usepackage[]{amsmath}
\usepackage{color}
\usepackage{xspace}
\usepackage{soul}

\usepackage{lineno}
\linenumbers

\newcommand{\msun}{\mathrm{M}_\odot}
\graphicspath{{figures/}}

\newcommand{\ud}{\mathrm{d}}

\def\lsim{ \lower .75ex \hbox{$\sim$} \llap{\raise .27ex \hbox{$<$}} }

\submitjournal{ApJ}

\shorttitle{satellites, stellar and dark matter halos in HSC}
\shortauthors{Wang et al.}
\graphicspath{{./}{figures/}}

\begin{document}


\title{The stellar mass in and around isolated central galaxies: connections to the total mass distribution through galaxy-galaxy lensing in the Hyper Suprime-Cam survey}

\correspondingauthor{Wenting Wang}
\email{wenting.wang@sjtu.edu.cn}


\author[0000-0002-5762-7571]{Wenting Wang}
\affiliation{Department of Astronomy, Shanghai Jiao Tong University, Shanghai 200240, China}
\affiliation{Shanghai Key Laboratory for Particle Physics and Cosmology, Shanghai 200240, China}
\author{Xiangchong Li}
\affiliation{Kavli IPMU (WPI), UTIAS, The University of Tokyo, Kashiwa, Chiba 277-8583, Japan}
\affiliation{Department of Physics, University of Tokyo, Tokyo 113-0033, Japan}
\author{Jingjing Shi}
\affiliation{Kavli IPMU (WPI), UTIAS, The University of Tokyo, Kashiwa, Chiba 277-8583, Japan}
\author[0000-0002-8010-6715]{Jiaxin Han}
\affiliation{Department of Astronomy, Shanghai Jiao Tong University, Shanghai 200240, China}
\affiliation{Shanghai Key Laboratory for Particle Physics and Cosmology, Shanghai 200240, China}
\affiliation{Kavli IPMU (WPI), UTIAS, The University of Tokyo, Kashiwa, Chiba 277-8583, Japan}
\author{Naoki Yasuda}
\affiliation{Kavli IPMU (WPI), UTIAS, The University of Tokyo, Kashiwa, Chiba 277-8583, Japan}
\author[0000-0002-4534-3125]{Yipeng Jing}
\affiliation{Department of Astronomy, Shanghai Jiao Tong University, Shanghai 200240, China}
\affiliation{Shanghai Key Laboratory for Particle Physics and Cosmology, Shanghai 200240, China}
\author{Surhud More}
\affiliation{The Inter-University Centre for Astronomy and Astrophysics, Post bag 4, Ganeshkhind, Pune 411007, India}
\author{Masahiro Takada}
\affiliation{Kavli IPMU (WPI), UTIAS, The University of Tokyo, Kashiwa, Chiba 277-8583, Japan}
\author{Hironao Miyatake}
\affiliation{Institute for Advanced Research, Nagoya University, Furo-cho, Nagoya 464-8601, Japan}
\affiliation{Division of Particle and Astrophysical Science, Graduate School of Science,Nagoya University, Furo-cho, Nagoya 464-8602, Japan}
\author{Atsushi J. Nishizawa}
\affiliation{Institute for Advanced Research, Nagoya University, Furo-cho, Nagoya 464-8601, Japan}




\begin{abstract}

Using photometrically selected galaxies from the Hyper Suprime-Cam (HSC) imaging survey, we measure 
the stellar mass density profiles for satellite galaxies as a function of the projected distance, $r_p$, to 
isolated central galaxies (ICGs) selected from SDSS/DR7 spectroscopic galaxies at $z\sim0.1$. By stacking HSC 
images, we also measure the projected stellar mass density profiles for ICGs and their stellar halos. The 
total mass distributions are further measured from HSC weak lensing signals. ICGs dominate within $\sim$0.15 
times the halo virial radius ($0.15R_{200}$). The stellar mass versus total mass fractions drop with the 
increase in $r_p$ up to $\sim0.15R_{200}$, beyond which the fractions are less than 1\% while stay almost 
constant, indicating the radial distribution of satellites trace dark matter. The integrated stellar mass 
locked in satellites is proportional to the virial mass of the host halo, $M_{200}$, for ICGs more massive 
than $10^{10.5}\msun$, i.e., $M_{\ast,\mathrm{sat}}\propto M_{200}$, whereas the scaling relation between 
the stellar mass of ICGs $+$ stellar halos and $M_{200}$ is close to $M_{\ast,\mathrm{ICG+diffuse}}\propto 
M_{200}^{1/2}$. Below $10^{10.5}\msun$, the change in $M_{200}$ is much slower with the decrease in $M_{\ast,
\mathrm{ICG+diffuse}}$. At fixed stellar mass, red ICGs are hosted by more massive dark matter halos and 
have more satellites. Interestingly, at $M_{200}\sim10^{12.7}\msun$, both $M_{\ast,\mathrm{sat}}$ and the 
fraction of stellar mass in satellites versus total stellar mass, $f_\mathrm{sat}$, tend to be marginally
higher around blue ICGs, perhaps implying the late formation of blue galaxies. $f_\mathrm{sat}$ increases 
with the increase in both $M_{\ast,\mathrm{ICG+diffuse}}$ and $M_{200}$, and scales more linearly with 
$M_{200}$. We provide best-fitting relations to $M_{200}$ versus $M_{\ast,\mathrm{ICG+diffuse}}$, 
$M_{\ast,\mathrm{sat}}$ or $M_{\ast,\mathrm{ICG+diffuse}}+M_{\ast,\mathrm{sat}}$, and to $f_\mathrm{sat}$ 
versus $M_{200}$ or $M_{\ast,\mathrm{ICG+diffuse}}$, for red and blue ICGs separately.

\end{abstract}



\section{Introduction} \label{sec:intro}

In the structure formation paradigm of $\Lambda$CDM, galaxies form by the cooling and
condensation of gas at centers of dark matter halos \citep{1978MNRAS.183..341W}. It is
usually believed that galaxy formation involves two phases, an early rapid formation of
``in-situ'' stars through gas cooling and a later phase of mass growth through accretion
of smaller satellite galaxies, which were originally central galaxies of smaller dark
matter halos. These smaller halos and galaxies, after falling into larger halos, become
the so-called subhalos and satellite galaxies. These satellites lose their stellar mass
due to tidal stripping. Stripped stellar materials typically lie in the outskirts of
central galaxies and are more metal poor than ``in-situ'' stars, which form the diffuse
light or the extended stellar halos around the central galaxies \citep[e.g.][]{2005ApJ...635..931B,
2010MNRAS.406..744C}.

Many observational studies on satellite galaxies and the faint diffuse stellar halos focused on the 
Local Group (LG) or the very nearby Universe because of the greater depth and detail with which 
nearby galaxies can be studied. Particular concerns have been paid to comparing $\mathrm{\Lambda} 
\text{CDM}$ predictions to the abundance and internal structure of dwarf spheroidal galaxies 
in and around the LG \citep[e.g.][]{1999ApJ...522...82K,1999ApJ...524L..19M,2009ApJ...696.2179K,
2011MNRAS.417.1260F,2011ApJ...738..102T,2011MNRAS.415L..40B} and to the low surface brightness 
stellar halos of disc galaxies similar to our Milky Way (MW) Galaxy \citep[e.g.][]{2016ApJ...830...62M,2017MNRAS.466.1491H,2020MNRAS.495.4570M,2021arXiv210309833K}. 
Such studies are limited by the fact that the nearby Universe contains limited number of galaxies, 
as considerable scatter is expected among systems of similar mass halos
\citep[e.g.][]{2015MNRAS.454..550G,2015MNRAS.452.3838C,2021ApJ...908..109C,2021ApJ...907...85M}.

Early studies of extra-galactic satellites in the more distant Universe often rely on spectroscopic 
surveys with redshift information to study their projected number density profiles to the central 
galaxies\citep[e.g.][]{1991ApJ...379L...1V,2005MNRAS.356.1045S,2006ApJ...647...86C}. 
Spectroscopic redshifts enable the direct discrimination between true satellites and background 
sources. However, satellites with redshift information are usually about only one to two magnitudes 
fainter than the central galaxy in most existing wide field spectroscopic surveys. To study 
smaller and fainter satellites, quite a few studies attempted to combine a redshift survey (for 
centrals) with a photometric survey (for satellites), which can reach several magnitudes fainter 
than pure spectroscopic surveys. Results can be achieved by stacking satellite counts around a 
large sample of central galaxies \citep[e.g.][]{1987MNRAS.229..621P,1994MNRAS.269..696L,
2011ApJ...734...88W,2011AJ....142...13L,2012MNRAS.424.2574W,2012MNRAS.427..428G,2012ApJ...751L...5T,
2012ApJ...760...16J,2013ApJ...772..146N,2014MNRAS.442.1363W,2015MNRAS.449.2576C,2016MNRAS.459.3998L}. 
Stacking not only helps to increase the signal-to-noise level, but also the foreground and 
background source contamination can be subtracted statistically.

Similarly, studies on low surface brightness stellar halos of distant extra-galactic 
systems often rely on stacking images of a large sample of galaxies with similar properties
\citep[e.g.][]{2004MNRAS.347..556Z,2005MNRAS.358..949Z,2011ApJ...731...89T,2014MNRAS.443.1433D,
2019MNRAS.487.1580W,2019ApJ...874..165Z}. This is because the surface brightness, $I$, of extended 
objects drops with distance, in a relationship with redshift, $z$, as $I\propto(1+z)^{-4}$. Hence 
for more distant galaxies, the surface brightness of their faint stellar halos can be only a few 
percent or even less than the sky background. This makes it more difficult to study distant stellar 
halos individually. Fortunately, the noise level of the stacked image can be significantly decreased 
with the increase in the number of input images used for stacking. It enables the detection of the 
averaged signals for extended stellar halos, which were once below the noise level of individual 
images. 

The observed abundance and properties of satellite galaxies, the total mass or emission in central 
galaxies and their diffuse stellar halos, and their connections to the host dark matter halos are 
important aspects to qualitatively test the theory of galaxy formation.
It has been recognized that more massive central galaxies have higher abundance of satellites and 
more dominant outer stellar halos \citep[e.g.][]{2011ApJ...734...88W,2011MNRAS.417..370G,2012ApJ...760...16J,
2014MNRAS.442.1363W,2014MNRAS.443.1433D,2016MNRAS.459.3998L}. In addition, there are more satellites 
and also more extended stellar halos around red passive centrals than blue centrals with the same stellar 
mass \citep[e.g.][]{2012ApJ...757....4P,2012MNRAS.424.2574W,2014MNRAS.443.1433D}. So far, observations 
are generally consistent with predictions by theoretical studies
\citep[e.g.][]{2007ApJ...666...20P,2010ApJ...725.2312O,2012MNRAS.425..641L,2014MNRAS.444..237P,
2016MNRAS.458.2371R,2018arXiv180810454K,2013MNRAS.434.3348C,2014ApJ...792..103K,2015MNRAS.451.2703C,
2016MNRAS.458.2371R}.

The halo mass versus stellar mass relation of central galaxies have been investigated in many 
previous studies, based on a few different approaches including abundance matching \citep[e.g.][]{
2010MNRAS.404.1111G}, Halo Occupation Distribution modelling \citep[e.g.][]{2010MNRAS.402.1796W,
2010ApJ...710..903M,2013MNRAS.428.3121M,2013MNRAS.431..648W,2013MNRAS.431..600W} or lensing measurements
\citep[e.g.][]{2012ApJ...744..159L,2015MNRAS.447..298H,Han15}. It has been generally recognized that 
the relation shows a transition around MW mass, above which halo mass changes more rapidly with stellar 
mass and below which halo mass changes very slowly with the decrease in stellar mass. However, many 
existing studies did not distinguish central galaxies by color upon determining the relations. 
In addition, the emissions in the outer stellar halos of massive elliptical galaxies are often failed 
to be detected in shallow surveys such as SDSS, resulting in under estimates in the total luminosity 
or stellar mass of central galaxies \citep[e.g.][]{2013ApJ...773...37H,2015MNRAS.454.4027D}.

Alternatively, many studies use satellite abundance as a proxy to the 
host halo mass \citep[e.g.][]{2012ApJ...757....4P,2012MNRAS.424.2574W,2013MNRAS.428..573S,2015ApJ...799..130R,
2019ApJ...881...74M}. Indeed, as have been proved by weak lensing measurements, red passive centrals not 
only have more satellites, but also they are hosted by more massive dark matter halos than blue centrals 
with the same stellar mass \citep{2016MNRAS.457.3200M}, indicating a tight relation between satellite 
abundance and host halo mass. Recently, \cite{2019arXiv191104507T} further suggested that the total 
luminosity of satellites brighter than a certain magnitude threshold can be used as a good proxy to 
the host halo mass.

In this paper, we at first re-investigate the role of satellites as proxies to their host halo mass, by 
counting and averaging photometric companions from the Hyper Suprime-Cam (HSC) deep imaging survey 
around spectroscopically identified isolated central galaxies selected from SDSS/DR7. We avoid the small 
scale regions close to the central galaxies, which are significantly affected by source deblending 
mistakes. We calculate the average projected stellar mass density profiles for satellites, and 
for red and blue central galaxies separately. The integrated stellar mass in these satellites will be 
directly compared with the best-constrained host halo mass through weak lensing signals.

In addition to satellites, the projected stellar mass density profiles for central galaxies and their
stellar halos will be measured and investigated in this paper as well, and for red and blue 
centrals separately. Based on the HSC imaging data products, \cite{2019MNRAS.487.1580W} (hereafter 
Paper I) have measured the PSF-corrected surface brightness profiles by stacking images of 
isolated central galaxies. In this study we follow the approach in Paper I but further convert the 
surface brightness profiles to projected stellar mass density profiles. The integrated stellar 
mass over the profile after stacking include the contribution from the faint outer stellar halos, 
which was not fully detectable in shallow surveys or before stacking. 

Combining the measurements of satellites, centrals and their stellar halos, we are able to 
investigate the radial distribution of total stellar mass versus dark matter. We also investigate 
the connection between satellites and central galaxies $+$ their diffuse stellar halos, including 
the fraction of stellar mass in satellites versus the total stellar mass, and the transition radius 
beyond which satellites dominate. This is the first attempt of measuring the projected stellar mass 
density profiles for both the diffuse stellar halos and satellite galaxies in the same paper.

We introduce our sample selection of halo central galaxies, the HSC photometric source 
catalog and imaging products, and the HSC shear catalog in Section~\ref{sec:data}. Our method of 
satellite counting and stacking, image stacking, PSF and $K$-corrections and lensing measurements 
are detailed in Section~\ref{sec:method}. Results are presented in Section~\ref{sec:result}.  
We conclude in the end (Section~\ref{sec:conclusion}).

We adopt as our fiducial cosmological model the first-year Planck cosmology \citep{2014A&A...571A..16P},
with values of the Hubble constant $H_0=67.3\mathrm{km s^{-1}/Mpc}$, the matter density $\Omega_m=0.315$
and the cosmological constant $\Omega_\Lambda=0.685$.

\section{data}
\label{sec:data}
\subsection{Isolated central galaxies}
\label{sec:isogal}

To identify a sample of galaxies with a high fraction of central galaxies in dark matter
halos (purity), we select galaxies that are the brightest within given projected and line-of-sight
distances. The parent sample used for this selection is the NYU Value Added Galaxy Catalog
\citep[NYU-VAGC;][]{2005AJ....129.2562B}, which is based on the spectroscopic Main galaxy sample
from the seventh data release of the Sloan Digital Sky Survey \citep[SDSS/DR7;][]{2009ApJS..182..543A}.
The sample includes galaxies in the redshift range of $z=0.001$ to $z\sim0.4$, which is flux
limited down to an apparent magnitude of 17.77 in SDSS $r$-band, with most of the objects
below redshift $z=0.25$.

Following \cite{2014MNRAS.443.1433D}, we at first exclude galaxies whose minor to major axis
ratios are smaller than 0.3, which are likely edge-on disc galaxies. \cite{2008MNRAS.388.1521D}
pointed out that the scattered light through the far wings of point spread function (PSF) from
edge-on disc galaxies can significantly contaminate the signal of stellar halos.

We require that galaxies are the brightest within the projected virial radius, $R_{200,\mathrm{AM}}$, 
of their host dark matter halos\footnote{$R_{200}$ is defined to be the radius within which the
average matter density is 200 times the mean critical density of the universe. Throughout this
paper, the virial mass is defined to be the total mass within $R_{200}$, denoted as $M_{200}$.}
and within three times the virial velocity along the line-of-sight direction. Moreover, these
galaxies should not be within the projected virial radius (also three times virial velocity
along the line-of-sight) of another brighter galaxy. The virial radius and velocity are derived
through the abundance matching formula of \cite{2010MNRAS.404.1111G}, and thus we use the index 
AM here to denote abundance matching. It has been demonstrated that the choice of three times 
virial velocity along the line-of-sight is a safe criterion that identifies all true companion 
galaxies \citep{2013MNRAS.428..573S}.

The SDSS spectroscopic galaxies suffer from the fiber collision effect that two fibers
cannot be placed closer than 55$\arcsec$. To avoid the case when a galaxy has a brighter
companion but this companion is not included in the SDSS spectroscopic sample, we use the
SDSS photometric catalog to make further selections. The photometric catalog is the
value-added Photoz2 catalog \citep{2009MNRAS.396.2379C} based on SDSS/DR7, which provides
photometric redshift probability distributions for SDSS galaxies. We further discard galaxies
that have a photometric companion, whose redshift is not available but is within the projected
separation of the given selection criterion, and its photoz probability distribution gives  a
larger than 10\% of probability that it shares the same redshift as the central galaxy.

We provide in the third to fifth columns of Table~\ref{tbl:icg} the numbers of all, red and blue isolated 
central galaxies. The color division is slightly stellar mass dependent (see \cite{2012MNRAS.424.2574W} 
for details). The total number of galaxies in each bin is larger than that in Paper I. This is because 
in Paper I we did not perform $K$-corrections, which accounts for the band shift due to cosmic 
expansion, and thus a narrower redshift range of $0.05<z<0.16$ was adopted to reduce the effect 
introduced by ignoring $K$-corrections. In this paper, we include $K$-corrections, and thus all 
ICGs over a wider redshift range are used. In Paper I, we investigated the purity and completeness 
of isolated central galaxies by using the mock galaxy catalog built from the Munich semi-analytical 
model of galaxy formation \citep{2011MNRAS.413..101G}. We found galaxies selected in this way have 
more than 85\% of being true halo central galaxies, and the completeness is above 90\%. Hereafter, 
we call our sample of isolated central galaxies in short as ICGs.

\subsection{HSC imaging data and photometric sources}
\label{sec:step}

\begin{table}
\caption{Average halo virial radius ($R_{200,\mathrm{mock}}$) and number ($N$) of all, red and blue isolated 
central galaxies in six stellar mass bins considered in this  study. $R_{200,\mathrm{mock}}$ is based on isolated 
central galaxies in a mock galaxy catalog built from the Munich semi-analytical model of galaxy formation
\citep{2011MNRAS.413..101G}, rather than using abundance matching.}
\begin{center}
\begin{tabular}{lcccc}\hline\hline
$\log M_*/\msun$  & \multicolumn{1}{c}{$R_{200,\mathrm{mock}}$ [kpc]} & \multicolumn{1}{c}{$N_\mathrm{galaxy}$} & \multicolumn{1}{c}{$N_\mathrm{red}$} & \multicolumn{1}{c}{$N_\mathrm{blue}$}\\ \hline
11.4-11.7  & 758.0  &  1008  &  964  & 44  \\
11.1-11.4   & 459.08 & 3576   & 3026  &  550  \\
10.8-11.1  & 288.16 & 6370  & 4345  & 2023  \\
10.5-10.8  & 214.80 & 6008  & 3142  &  2861 \\
10.2-10.5  & 173.18 & 3771 & 1353  & 2418  \\
9.9-10.2  & 121.1 & 3677  &  700 &  2977 \\
\hline
\label{tbl:icg}
\end{tabular}
\end{center}
\end{table}

HSC-SSP \citep{2018PASJ...70S...4A} is based on the new prime-focus camera, the Hyper Suprime-Cam
\citep{2012SPIE.8446E..0ZM,2018PASJ...70S...1M,2018PASJ...70S...2K,2018PASJ...70S...3F} on the
8.2-m Subaru telescope. It aims for a wide field of 1,400 deg$^2$ with a depth of $r\sim26$, a
deep field of 26 deg$^2$ with a depth of $r\sim27$ and an ultra-deep field of 3.5 deg$^2$ with
one magnitude fainter. In this paper we focus on the wide field data. HSC photometry covers five
bands (HSC-$grizy$). The transmission curves and wavelength ranges for HSC $gri$-bands are almost
the same as those of SDSS \citep{2018PASJ...70...66K}.

HSC-SSP data is processed using the HSC pipeline, which is a specialized version of the
LSST \citep{2010SPIE.7740E..15A,2017ASPC..512..279J} pipeline code. Details about the HSC
pipeline are available in the pipeline paper \citep{2018PASJ...70S...5B}, and here we only
introduce the main data reduction steps and the corresponding data products of HSC.

In the context of HSC, a single exposure is called one ``visit'' with a unique ``visit''
number. The same sky field is ``visited'' multiple times. The HSC pipeline involves four
main steps: (1) processing of single exposure (one visit) image (2) joint astrometric
and photometric calibration (3) image coaddition and (4) coadd measurement.

In the first step, bias, flat field and dark flow are corrected for. Bad pixels, pixels
hit by cosmic rays and saturated pixels are masked and interpolated. The pipeline 
first performs an initial background subtraction before source detection. Detected 
sources are matched to external reference catalogs in order to calibrate the zero point 
and a gnomonic world coordinate system for each CCD. After galaxies and blended objects 
are filtered out, the sky background is estimated and subtracted again, based on 
blank pixels that do not contain any detection. A secure sample of stars are used to 
construct the central PSF model\footnote{The pipeline mainly returns the PSF model within 
$\sim3\arcsec$, which is the part dominated by the atmosphere turbulence.}. The outputs 
of this step are called Calexp images (means ``calibrated exposure''). They are given on 
individual exposure basis.

In the second joint calibration step, the astrometric and photometric calibrations are refined
by requiring that the same source appearing on different locations of the focal plane during
different visits should give consistent positions and fluxes. Readers can find details in
\cite{2018PASJ...70S...5B}. The joint calibration improves the accuracy of both astrometry
and photometry.

In the third step, the HSC pipeline resamples images to the pre-defined output skymap. It
involves resampling both the single exposure images and the PSF model \citep{2011PASP..123..596J}
using a 3rd (for internal releases earlier than and also including S18a) or 5th-order (For
S19a and later releases) Lanczos kernel. Resampled images of different visits are then
combined together (coaddition). Images produced through coaddition are called coadd images.

In the last step, objects are detected, deblended and measured from the coadd images. A
maximum-likelihood detection algorithm is run independently for each band. Background is
estimated and subtracted once again. The detected footprints and peaks of sources are
merged across different bands to maintain detections that are consistent over different
bands and to remove spurious peaks. These detected peaks are deblended and a full suite
of source measurement algorithm is run on all objects, yielding independent measurements
of source positions and other properties in each band. A reference band of detection is
then defined for each object based on both the signal-to-noise ratio (SNR) and the purpose 
of maximizing the number of objects in the reference band. Finally, the measurement of 
sources are run again with the position and shape parameters fixed to the values in the 
reference band to achieve the ``forced'' measurement. The forced measurement brings 
consistency across bands and enables computing object colors using the magnitude
 difference in different bands.

The faint stellar halo to be studied in this paper can be less than a few percent of the
sky background, and thus it is crucial to properly model and subtract the sky background.
The HSC internal data releases S15, S16 and S17 used a 6th-order Chebyshev polynomial to fit
individual CCD images to model the local sky background and instrumental features\footnote{The
first public data release is similar to the internal S15b data release.}. It over-subtracts the
light around bright sources and leaves a dark ring structure around bright galaxies. The
over-subtraction is mainly caused by the scale of the background model (or order of the
polynomial fitting) and unmasked outskirts of bright objects. It is difficult to know how
extended objects are before coaddition.

The internal S18a data release\footnote{The second public data release is almost identical to the
internal S18a release} implements a significantly improved global background subtraction approach
\citep{2019PASJ...71..114A}. An empirical background model crossing all CCDs are used to model the
sky background, meaning that discontinuities at CCD edges are avoided. A scaled ``frame'', which
is the mean response of the instrument to the sky for a particular filter, is used to correct for
static instrumental features that have a smaller scale than the empirical background model. As have
been shown in \cite{2019PASJ...71..114A} and \cite{2019MNRAS.487.1580W}, artificial instrumental
features over the focal plane have been successfully removed following the use of the S18a data,
leaving uniform image backgrounds. The S18a release also adopts a larger scale of about 1000 pixels
to model the sky background, which minimizes the problem of oversubtraction.

While the global background subtraction approach has helped to avoid
over-subtracting the extended outskirts of galaxies, it introduces new
issues due to sky background residuals. One notable issue is that cModel fluxes
are over-estimated for some of the faint sources, and occasionally even a fake
detection has a large flux. Thus for the S19a data release, the pipeline performs 
the global background subtraction as S18a when processing single exposures, but 
it enables a local background subtraction upon coadd image processing for detailed 
measurements (e.g., galaxy shape and photometric redshift estimations). The issues 
due to sky background residuals have been mitigated in S19a though not completely 
eliminated.

In our analysis throughout the main text of this paper, we use coadd images of the S18a internal data
release to stack galaxy images and calculate the projected stellar mass density profiles of central
galaxies and their stellar halos. In addition, we use primary photometric sources which are classified
as extended from the S19a internal data release to calculate the radial distribution of satellite
galaxies. We exclude sources with any of the following flags set as true in $g$, $r$ and $i$-bands:
bad, crcenter, saturated, edge, interpolatedcenter or suspectcenter. The S19a bright star masks are
created based on stars from {\it Gaia} DR2, and we have excluded sources within the ghost, halo and
blooming masks of bright stars in $i$-band. We also limit to footprints reaching full depth in $g$
and $r$-bands. As have been discussed by \cite{2018PASJ...70S...8A} and \cite{2021MNRAS.500.3776W},
the completeness of photometric sources in HSC is very close to 1 at $r\sim25$, and thus we adopt a
flux limit of $r<25$. The total area is a bit more than 450 square degrees.

\subsection{HSC shear catalog}
We use the HSC second-year shape catalog (Li et al., in prep.) produced from
the $i$-band wide field of HSC S19a internal data release (Aihara et al., in prep.). 
The catalog covers an area of $433.48 ~\rm{deg}^2$ with a mean seeing of $0.59\arcsec$. 
With conservative galaxy selection criteria, the raw galaxy source number density 
are $22.94 / \rm{arcmin}^{2}$.

The shapes of galaxies are estimated with the re-Gaussianization PSF correction
method \citep{Regaussianization}. The output of the re-Gaussianization
estimator is the galaxy ellipticity:
\begin{equation}
    (e_1,e_2)=\frac{1-(b/a)^2}{1+(b/a)^2} (\cos 2\phi,\sin 2\phi),
\end{equation}
where $b/a$ is the axis ratio, and $\phi$ is the position angle of the major
axis with respect to the equatorial coordinate system.

The shapes are calibrated with realistic image simulations downgrading the
galaxy images from Hubble Space Telescope \citep{HST-ACSpipe} to the HSC galaxy
images \citep{HSC1-GREAT3Sim}. The calibration removes the galaxy
property-dependent (galaxy resolution, galaxy SNR, and galaxy redshift) shear
estimation bias (i.e., multiplicative bias and additive bias).
The multiplicative bias and additive bias for a galaxy ensemble are:
\begin{equation}
\begin{split}
    \hat{m}&=\frac{\sum_i w_i m_i}{\sum_i w_i},\\
    \hat{c}_{1,2}&=\frac{\sum_i w_i a_i e^{\rm{PSF}}_{1,2;i}}{\sum_i w_i},
\end{split}
\end{equation}
respectively. Here, $i$ refers to the galaxy index. $w_i$, $m_i$ and $a_i$ are
the galaxy shape weight, multiplicative bias and fractional additive bias for
galaxy $i$. The galaxy shape weight is defined as
\begin{equation}
    w=\frac{1}{\sigma_e^2+e_{\rm{rms}}^2},
\end{equation}
where $e_{\rm{rms}}$ is the root-mean-square ($\texttt{RMS}$) of intrinsic
ellipticity per component, and $\sigma_e$ refers to the shape measurement error
per component due to photon noise. $e_{\rm{rms}}$ and $\sigma_e$ are modeled
and estimated for each galaxy using the image simulation. The calibrated shear
estimation for the galaxy ensemble is:
\begin{equation}
    \hat{g}_{1,2}=\frac{\sum_i w_i e_{1,2;i}}{2 \mathcal{R} (1+\hat{m})\sum_i w_i}
    -\frac{\hat{c}_{1,2}}{1+\hat{m}}.
    \label{eqn:shear}
\end{equation}
Here $\mathcal{R}$ is the shear responsivity, defined as the response of the ensemble
averaged ellipticity to a small shear \citep{2002AJ....123..583B}.

\section{Method}
\label{sec:method}

In the following, we introduce in Sec.~\ref{sec:satcount} the satellite counting and background subtraction
methodologies. To calculate the stellar mass distribution of ICGs and their diffuse stellar halos, we at 
first stack galaxy images to obtain the PSF-corrected surface brightness profiles (Section~\ref{sec:SB}). 
$K$-corrections are then achieved for each individual galaxy (Section~\ref{sec:kcorr}). In the end, the 
surface brightness profiles are converted to the projected stellar mass density profiles based on the
$K$-corrected and PSF free color profiles (Section~\ref{sec:masscalc}). We introduce how the differential 
total mass density profiles are calculated from weak lensing signals in Section~\ref{sec:lensing}.

\subsection{Satellite counts and background subtraction}
\label{sec:satcount}

To calculate the projected density profiles of stellar mass in resolved satellites around ICGs, we make use
of the HSC source catalog, which is flux limited down to the $r$-band apparent magnitude of $r\sim25$.

We follow the method of \cite{2012MNRAS.424.2574W} and \cite{2014MNRAS.442.1363W} to make satellite counts.
Around each ICG, we count all companions in projected radial bins, with the projected distances to ICGs
computed from the angular separation and the redshift of the ICG. This includes both true satellites
and fore/background contaminations. For each companion, the $K$-correction formula of \cite{2010PASP..122.1258W}
is applied by using its apparent color and also assuming it is at the same redshift as the ICG. Further 
after distance modulus correction using the redshift of the central ICG, we obtain absolute magnitudes
and rest-frame colors. A conservative red end cut of $^{0.1}(g-r)<0.065\log_{10}M_\ast/\msun+0.35$ is then
made to the rest-frame colors of companions to eliminate the population of background sources which are too
red to be at the same redshifts of ICGs, and hence increase the signal. To estimate the stellar-mass-to-light
ratios, we adopt a Gaussian Process Regression (GPR) fitting procedure, which estimates the stellar mass 
of each companion through its color and will be introduced in detail in Sec.~\ref{sec:masscalc}.

Weighting companion number counts in different radial bins by their stellar mass, we obtain their projected
stellar mass density profiles around each ICG, which are based on companions with $\log_{10}M_{\ast,
\mathrm{ICG}}-3<\log_{10}M_{\ast,\mathrm{sat}}<\log_{10}M_{\ast,\mathrm{ICG}}$. To ensure the completeness 
of satellites given the HSC flux limit of $r<25$, we allow a particular central ICG to contribute counts only 
if the stellar mass corresponding to $r=25$ at the redshift of the ICG and lying on the red envelope of the 
intrinsic color distribution\footnote{To convert $r=25$ to an absolute magnitude and then to a stellar mass 
limit, we have to assume a mass-to-light ratio, which is the highest given the reddest intrinsic color allowed 
at the corresponding redshift of the central ICG. This gives the highest hence the safest limit in stellar mass.} 
is smaller than $\log_{10}M_{\ast,\mathrm{ICG}}-3$. Counts around all ICGs in the same stellar mass bin are 
cumulated  and averaged to give the final profiles.

To subtract fore/background sources, we repeat exactly the same steps with a sample of random points, which are
assigned the same redshift and stellar mass distributions as ICGs. Centered on these random points, we calculate 
a suite of random profiles. The random profiles are subtracted from the profiles centered on real ICGs, to obtain 
the projected stellar mass density profiles of real satellite galaxies.

Throughout the analysis of this paper, we consider the radial range, which is within 0.1 times the halo virial 
radius, $0.1R_{200}$, as have been significantly affected by source deblending issues. We provide details about 
how this inner radius cut is determined in Appendix~\ref{app:deblending}. In the following sections, results over 
the entire radial range will still be presented, but the measured profiles of satellite galaxies within $0.1R_{200}$
should be avoided for any scientific inferences.

\subsection{Surface brightness profiles of ICGs and stellar halos}
\label{sec:SB}

\begin{figure}
	\includegraphics[width=0.49\textwidth]{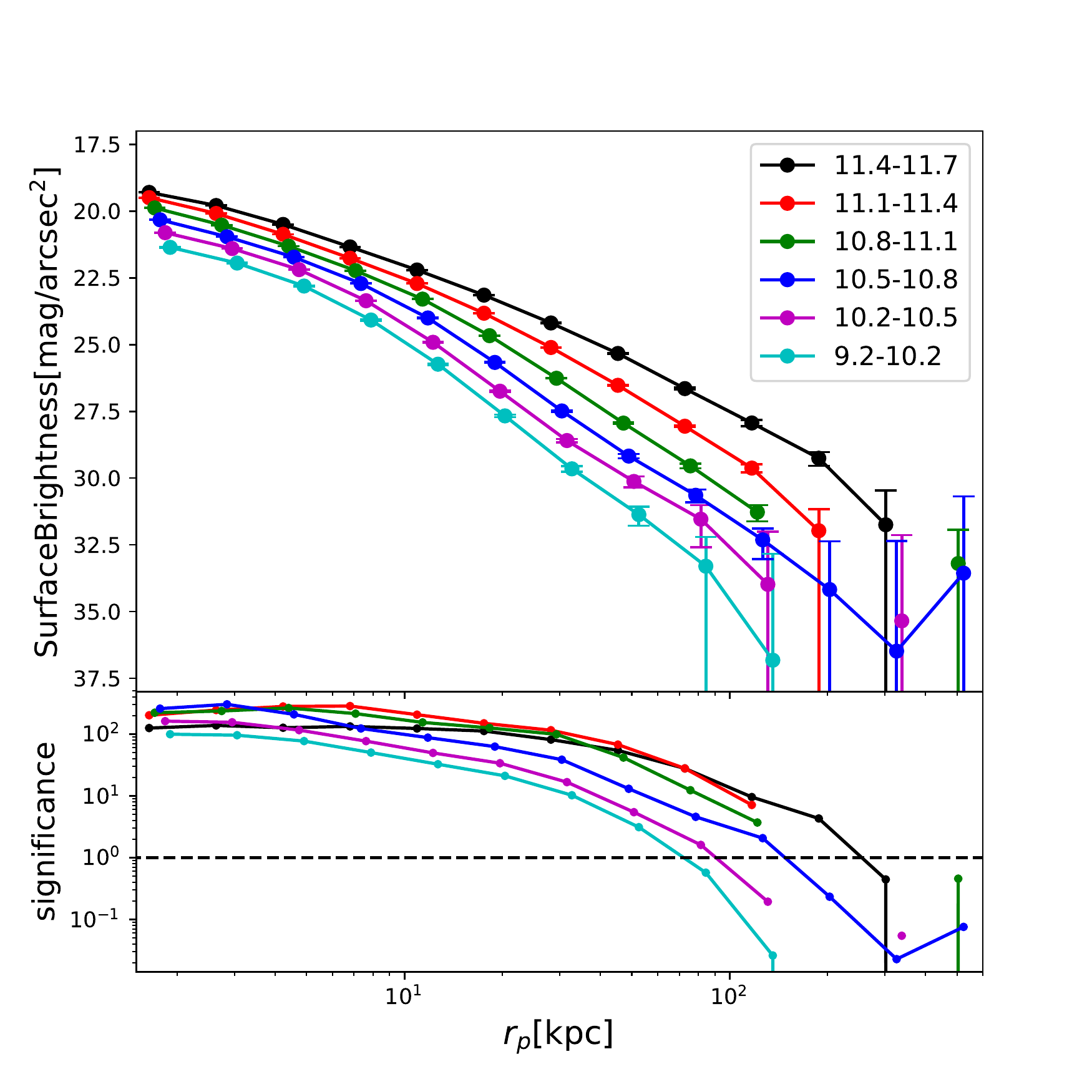}
	\caption{{\bf Upper panel:} Averaged surface brightness profiles for ICGs $+$ their
	stellar halos. ICGs are grouped into a few stellar mass bins, as indicated by the
	legend in log stellar mass. All ICGs in the redshift range of $0.01<z<0.3$ are used.
	No $K$-correction or PSF-correction have been applied. Errorbars are 1-$\sigma$
	scatters of 100 boot-strapped subsamples. {\bf Lower panel:} Measured surface brightness
	profiles divided by their 1-$\sigma$ errors.
	}
	\label{fig:profzall}
\end{figure}

The readers can find details about how we process galaxy images and calculate the surface
brightness profiles of ICGs and their outer stellar halos in Paper I. In this subsection,
we introduce our main steps.

Centered on each ICG, we extract its image cutout with edge length of $2\times1.3$ times the
virial radius, $R_{200,\mathrm{mock}}$. Here the values of $R_{200,\mathrm{mock}}$ for ICGs 
in a few different ranges of stellar mass are provided in the second column of Table~\ref{tbl:icg}, 
which are estimated from ICGs in the mock galaxy catalog of \cite{2011MNRAS.413..101G}. The 
physical scales are converted to angular sizes according to the redshifts of the ICGs. Each 
image cutout is divided by the zero point flux. Bad pixels such as those which are saturated, 
close to CCD edges, outside the footprint with available data, hit by cosmic rays and so on, 
are masked. We correct the effect of  "cosmic dimming" by multiplying the image cutout of
each ICG by $(1+z)^4$. The cutouts are resampled to ensure the same number of pixels within 
$R_{200,\mathrm{mock}}$. Each pixel has a fixed physical scale of 0.8~kpc.

To look at the ICGs and their diffuse stellar halos, all other companion sources, including
physically associated satellites and fore/background sources, are detected and masked.
Here we choose a combination of 0.5, 1.5, 2 and 3 times the background noise levels as detection
thresholds and successively apply them to the image cutouts. Note when adopting 0.5 times the
background noise as the threshold, we only use footprints which are also associated with sources
detected by 1.5 times the background noise level as well. This is to avoid faked detections
below the noise level. The readers can check Paper I for more detailed comparisons on different
choices of source detection and masking thresholds.

We stack images for ICGs with similar properties (e.g. in the same stellar mass bin of Table~\ref{tbl:icg}
and/or having similar colors). Stacking helps us to go beyond the noise level of individual images. For
each pixel in the output image plane, we at first clip corresponding pixels from all input images (masked
pixels are not included) by discarding 10\% pixels near the two ends of the distribution tail. We have
checked that varying this fraction between 1\% and 10\% does not bias the stacked surface brightness
profiles.

In the end, to account for any residual sky background, we repeat the same steps for image cutouts
centered on a sample of random points within the HSC footprint, and the random stacks are subtracted
from the stacks of real galaxies. These random points are assigned the same redshift and stellar mass
distributions as real galaxies. As have been shown in Paper I, the random stacks look ideally uniform
and show flat surface brightness profiles which agree very well with the large-scale surface brightness
profiles centered on real galaxies. In addition, the random stacks can account for incomplete masking
of fore/background companion sources.

Many previous studies tried to align galaxy images along their major axis before stacking and calculate
the surface brightness profiles based on elliptical isophotal contours, which are reported as functions
of the semi-major axis lengths. In our analysis, we choose not to rotate galaxy images. So basically
our stacked images are circularly averaged. As we have discussed in Paper I, because we have excluded
extremely edge-on galaxies, circularly averaged profiles show only slightly steeper color profiles than
major-axis aligned and elliptically averaged profiles if the color gradient is negative, and the surface
brightness profiles are not significantly affected.

The averaged 1-dimensional surface brightness profiles can be obtained from stacked images, which 
are shown by Figure~\ref{fig:profzall} for ICGs in a few different stellar mass bins. Slightly better 
signals are obtained thanks to the larger sample size than Paper I, especially for the most and least 
massive stellar mass bins. In addition to the final stacked images and surface brightness profiles, 
the image cutouts and profiles of individual ICGs processed in this step are saved as well.

After achieving measurements for the 1-dimensional surface brightness profiles, PSF-corrections 
are made to the profiles in HSC $g$, $r$ and $i$-bands. The corrections are made according to  
the extended PSF wings measured in Paper I and by fitting PSF-convolved model profiles to the 
measured surface brightness profiles, in order to estimate the PSF-contamination fraction as a function 
of projected distance to the center of galaxies. The details are provided in Appendix~\ref{app:psf}. 
Explicitly, the fraction increases with the increase in projected distance and decrease in stellar 
mass of ICGs. Interestingly, the fraction is higher around blue late-type galaxies than red early-type 
galaxies with the same stellar mass. This is mainly because blue galaxies are more extended, and their 
outer stellar halos are fainter. As a result, PSF tends to scatter more light from the central parts 
of galaxies to contaminate the signals of the true outer stellar halos.

\subsection{$K$-corrections}
\label{sec:kcorr}

\begin{figure}
\includegraphics[width=0.49\textwidth]{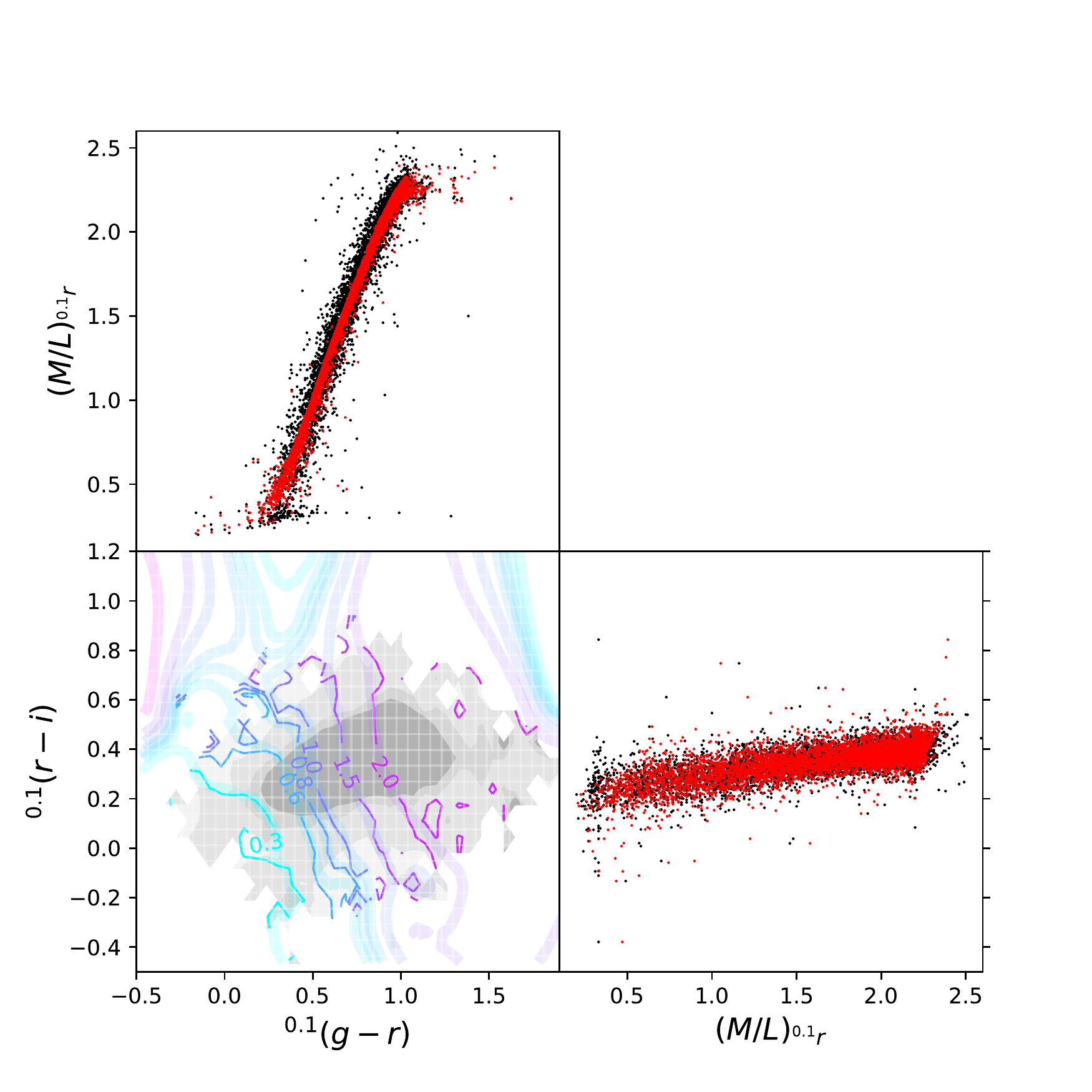}
\caption{Gaussian Process Regression (GPR) is adopted to recover the $r$-band stellar-mass-to-light
ratios ($M_\ast/L_r$) of SDSS spectroscopic Main galaxies from their $^{0.1}(g-r)$ and $^{0.1}(r-i)$
colors. The black dots are original values of a randomly picked up 10\% subsample used for testing
and validation, in which $M_\ast/L_r$ was estimated from the $K$-corrected colors by fitting a stellar
population synthesis model \citep{2007AJ....133..734B} assuming a Chabrier (2003) initial mass function
to the SDSS photometry. Red dots are recovered $M_\ast/L_r$ through GPR. Grey shaded region in the
bottom left panel demonstrate the number density distributions of SDSS Main galaxies, while colored
contours show the recovered $M_\ast/L_r$ through GPR.
 }
\label{fig:gpr}
\end{figure}

After achieving PSF corrections for individual ICGs in $g$, $r$ and $i$ filters, we conduct $K$-corrections
using the PSF-free color profiles and the redshift of each individual ICG. The empirical $K$-correction formula
provided by \cite{2010PASP..122.1258W} is adopted for the correction. Note the individual color profiles can
be noisy, and when estimates of the color is not available due to negative flux, we do linear interpolations
to obtain the color from neighboring radial bins with non-negative flux values.  After including $K$-corrections,
we calculate the rest-frame color and surface brightness profile for each individual ICG and its stellar halo,
in $g$, $r$ and $i$-bands.

To test the robustness of $K$-corrections, Figure~\ref{fig:colorevolve} in Appendix~\ref{app:kcorr} shows the
$g-r$ color profiles of ICGs split into two redshift bins ($z<0.1$ and $z>0.1$) before and after $K$-corrections.
The $g-r$ colors of massive galaxies at different redshifts after the $K$-correction agree well with each other,
and there are no obvious indications of failures in $K$-corrections. We have also checked the color profiles 
for individual ICGs before and after $K$-corrections, and the trends are all self-consistent.

\subsection{Stellar mass profiles of ICGs and their stellar halos}
\label{sec:masscalc}

Here we introduce how the surface brightness profiles of ICGs $+$ their stellar halos are converted to the projected
stellar mass density profiles. In a previous study, \cite{2018MNRAS.475.3348H} obtained the stellar mass profiles for
individual massive galaxies assuming that the massive galaxies can be well described by an average stellar-mass-to-light
ratio ($M_\ast/L$). \cite{2018MNRAS.475.3348H} achieved SED fitting and $K$-correction using the five-band HSC cModel
magnitudes. The approach of \cite{2018MNRAS.475.3348H} is more reasonable for their massive galaxies, which show shallow
color gradients. In our analysis, we probe a much broader range in stellar mass, and smaller galaxies can have much
steeper color gradients (see Paper I). Thus we use the radius-dependent color profiles to infer $M_\ast/L$ at different
radius.

To estimate the $r$-band stellar-mass-to-light ratios ($M_\ast/L_r$), we adopt a machine leaning approach of
Gaussian Process Regression (GPR) to fit the average dependence of $M_\ast/L_r$ on rest-frame $^{0.1}(g-r)$
and $^{0.1}(r-i)$ colors. SDSS spectroscopic Main galaxies are used for the training. GPR is a method which
fits a Gaussian process (GP) to data points. It allows for fitting arbitrary data points in multiple dimensions
without assuming any functional form. In this study, we use GPR as a flexible non-parametric smooth fit to
$M_\ast/L_r$ versus $^{0.1}(g-r)$ and $^{0.1}(r-i)$ colors. A thorough understanding of GPR is not essential
for this paper, as long as the readers can recognize GPR to be a non-parametric interpolation method.
The readers can find more details about GPR from the Appendix of \cite{2019MNRAS.482.1900H}.

We randomly pick up a 90\% subsample of SDSS spectroscopic Main galaxies for training. The remaining 10\% subsample
is used for validation. The original values of $M_\ast/L_r$ versus colors of the validating subsample are demonstrated
as black points in Figure~\ref{fig:gpr}, while red points are the predictions through GRP. It is very encouraging to
see that the red points trace well the black ones. In addition, we can see from the bottom left panel that 
$M_\ast/L_r$ shows a strong dependence on $^{0.1}(g-r)$, while the dependence on $^{0.1}(r-i)$ is still present 
though much weaker compared with that on $^{0.1}(g-r)$. Note our GPR only fits the average relation between 
$M_\ast/L_r$ and the colors. The prominent dependence of $M_\ast/L_r$ on $^{0.1}(g-r)$ appears sufficient to explain 
the scatter in $M_\ast/L_r$ at fixed  $^{0.1}(r-i)$ in the lower right panel. The remaining scatter around the red 
points at fixed $^{0.1}(g-r)$ in the top panel are those attributed to noises in our fitting.

In Appendix~\ref{app:stellarmasstest}, we also provide a scatter plot showing the original SDSS stellar mass
versus the predicted stellar mass. The recovered stellar mass agrees very well with the original value on
average, with very small biases.

We then use GPR to predict $M_\ast/L_r$ at different projected radius for ICGs $+$ their stellar halos. We use
the actual color profiles instead of a single aperture color for the ICG and its extended stellar halo as a 
whole. In the end, all the individual profiles are averaged, with 3-$\sigma$ clippings, to obtain the averaged 
projected stellar mass density profiles for ICGs grouped by similar stellar mass or color.

In Appendix~\ref{app:stellarmasstest}, we show a comparison of the stellar mass versus halo mass relation, where
the stellar mass is determined in a few different ways, including i) the integrated stellar mass over the measured 
stellar mass density profiles of ICGs $+$ their stellar halos, which is based on the radius dependent color profiles 
to infer $M_\ast/L_r$; ii) the integrated stellar mass over the stellar mass density profiles, but using a single 
aperture color to infer $M_\ast/L_r$ or iii) the integrated stellar mass similar to i) but without PSF deconvolutions; 
iv) the original stellar mass from SDSS (NYU-VAGC), which was obtained by SED fitting to the galaxy colors.

At the massive end, the integrated stellar mass is larger than the original stellar mass from SDSS. This is
expected, because SDSS is shallower than HSC, and by stacking galaxy images, we are able to push beyond the
noise level of individual images and detect more of the faint extended stellar halos. This is particularly
true for massive galaxies, which accreted more material and substructures to build their stellar halos. 
In order to properly measure the luminosity and stellar mass densities at the massive end, it is very 
important to carefully account for the stellar mass in the outer diffuse stellar halos \citep[e.g.][]{
2013ApJ...773...37H} with method iii above. Moreover, for smaller galaxies with steeper color gradients, 
we found, if applying the same aperture color to the whole radial range, the integrated stellar mass can be 
slightly over-estimated (see Appendix~\ref{app:stellarmasstest} for more information). Note, throughout this 
paper, we still adopt the stellar mass from SDSS to split ICGs into bins of stellar mass.

\subsection{Measuring the differential density profiles from weak lensing signals}
\label{sec:lensing}

In this study we follow the method described in \cite{2018PASJ...70S..25M} to calculate the mean differential 
projected density profiles for the total mass distribution around ICGs. In this subsection, we only summarize 
the main points.

The mean differential projected density profile, $\Delta \Sigma(r_p)$, is defined as the difference between
the mean surface density enclosed by projected radius $r_p$ (denoted $\bar{\Sigma}(<r_p)$ and the mean surface
density at that radius (denoted $\Sigma(r_p)$). The quantity $\Delta \Sigma(r_p)$ can be related to the mean
tangential shear of background source galaxies, $\gamma_t$, and the lensing critical density, $\Sigma_c$,

\begin{equation}
\Delta \Sigma=\langle \gamma_t \rangle \Sigma_c,
\label{eqn:diffdensp}
\end{equation}
where $\Sigma_c$ is defined as
\begin{equation}
 \Sigma_c=\frac{c^2}{4\pi G}\frac{D_s}{D_{ls} D_l}.
\end{equation}
$D_l$ and $D_s$ refer to the angular diameter distances of lens and source, respectively, and $D_{ls}$ is the
angular diameter distance between lens and source. Note throughout this paper, we use physical separations in
our analysis rather than comoving separations.

The mean tangential shear can be related to the directly measurable mean tangential ellipticity, $e_t$, of source
galaxies, and the two differing by a factor of twice the shear responsivity, $\mathcal{R}$. Thus $\Delta \Sigma$
is calculated based on Equation~\ref{eqn:shear}, in which $e_{1,2}$ is replaced by $e_t \Sigma_c$. In our 
calculation, we additionally use $1/\Sigma_c^2$ as weights to optimize the SNR.

For each lens galaxy with redshift $z_l$, source galaxies are selected as those with photometric
redshifts greater than $z_l+0.2$. The photometric redshifts are computed with dNNz (deep Neural Network 
photo-z). The readers can find more details about HSC photometric redshifts in \cite{2020arXiv200301511N}. The 
typical errors in HSC photometric redshifts are about 0.1, which can result in a typical underestimate in the 
weak lensing measured mass of only $\sim 0.02$ dex (\citealp{2015MNRAS.446.1356H}, see also \citealp{2012MNRAS.420.3240N}) 
at low redshifts. Besides, we have also tested alternative choices of $z_l+0.3$ and $z_l+0.4$, and the conclusions 
of this paper are almost the same. Thus photometric redshift errors are unlikely to have significantly affected our 
conclusions.


\section{Results}
\label{sec:result}

\subsection{Projected stellar mass density profiles of satellites, ICGs and their stellar halos}

\begin{figure}
\includegraphics[width=0.49\textwidth]{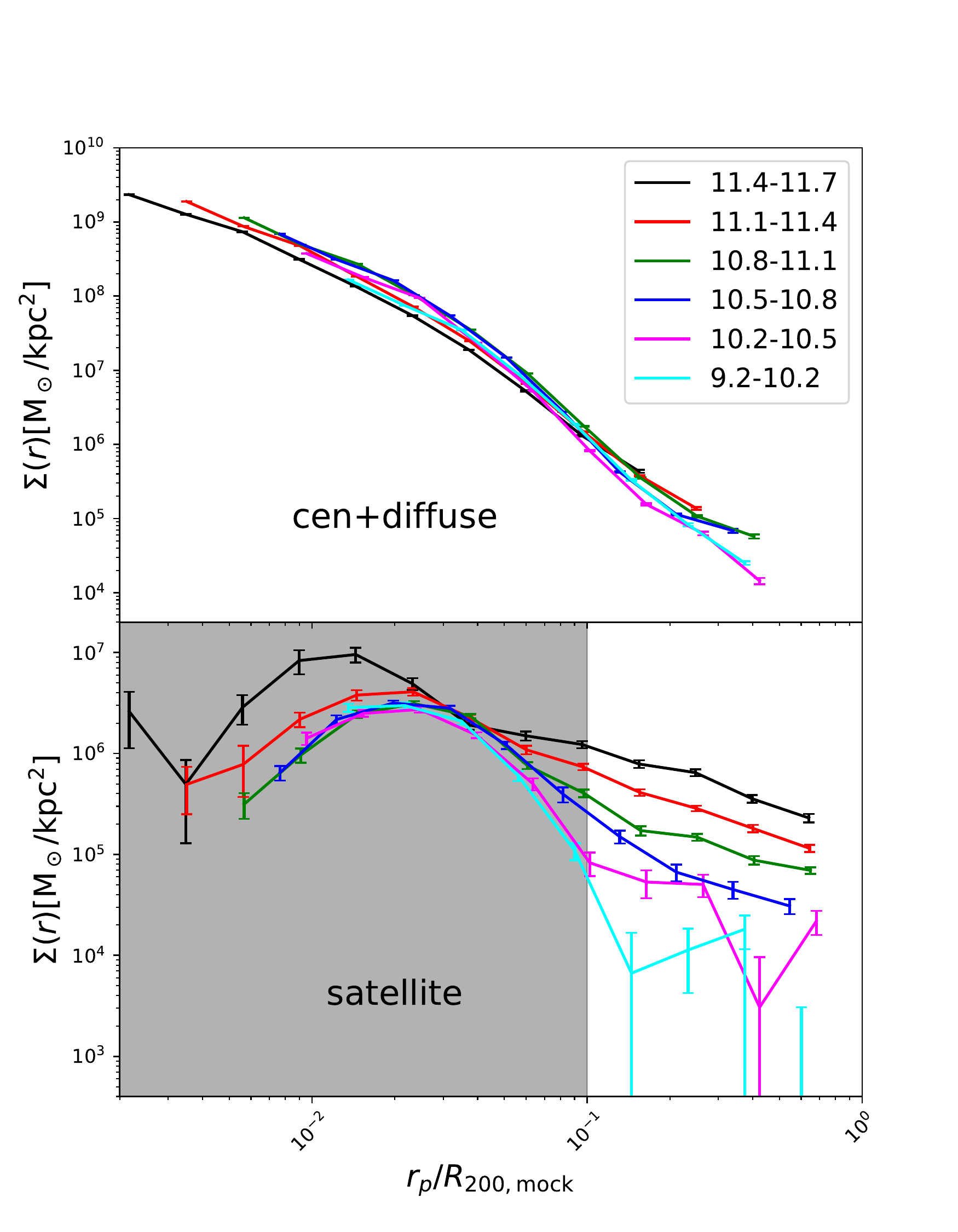}%
\caption{Projected stellar mass density profiles of ICGs $+$ their stellar halos (top) and of satellites (bottom),
centered on ICGs in different stellar mass bins (stellar mass taken from SDSS). The log stellar mass ranges are
shown by the legend. Errorbars are based on the 1-$\sigma$ scatters of 100 boot-strap subsamples. The gray shaded 
region in the bottom panel marks the radial range where the satellite profiles are significantly affected by 
deblending issues.
}
\label{fig:profmassall}
\end{figure}

The projected stellar mass density profiles are presented in Figure~\ref{fig:profmassall}, for ICGs grouped into a
few different stellar mass bins. {\it Hereafter, we call the projected stellar mass density profiles in short as
profiles.} The profiles of ICGs $+$ their extended stellar halos and of satellites are shown in the top and bottom
panels, respectively. The projected distance, $r_p$, has been scaled by the halo virial radius (see Table~\ref{tbl:icg}), 
$R_{200,\mathrm{mock}}$, which will be the convention adopted in most of the figures in following sections. In Paper 
I, we reported that the surface brightness profiles of ICGs and their stellar halos are very similar to each other 
after scaled by $R_{200,\mathrm{mock}}$, which is particularly true over the stellar mass range of $10.2<\log_{10}
M_\ast/\msun<11.1$. After scaled by $R_{200,\mathrm{mock}}$, the stellar mass density profiles after PSF-deconvolution 
also become more similar over the stellar mass range of $9.2<\log_{10}M_\ast/\msun<11.7$, as revealed by the top 
plot of Figure~\ref{fig:profmassall}.

The profiles of satellites, on the other hand, show very different trends. As demonstrated by the bottom panel
of Figure~\ref{fig:profmassall}, the profiles centered on more massive ICGs have higher amplitudes after scaling
$r_p$ by $R_{200,\mathrm{mock}}$ and at $r_p/R_{200,\mathrm{mock}}>0.1$, indicating the scaling relation between 
the stellar mass in satellites and the host halo mass must be different, which we will discuss later in Section~\ref{sec:scaling}. 
The profiles of satellites are also more flattened and extended than those of the central ICGs $+$ their stellar halos.

\begin{figure*}
\includegraphics[width=0.9\textwidth]{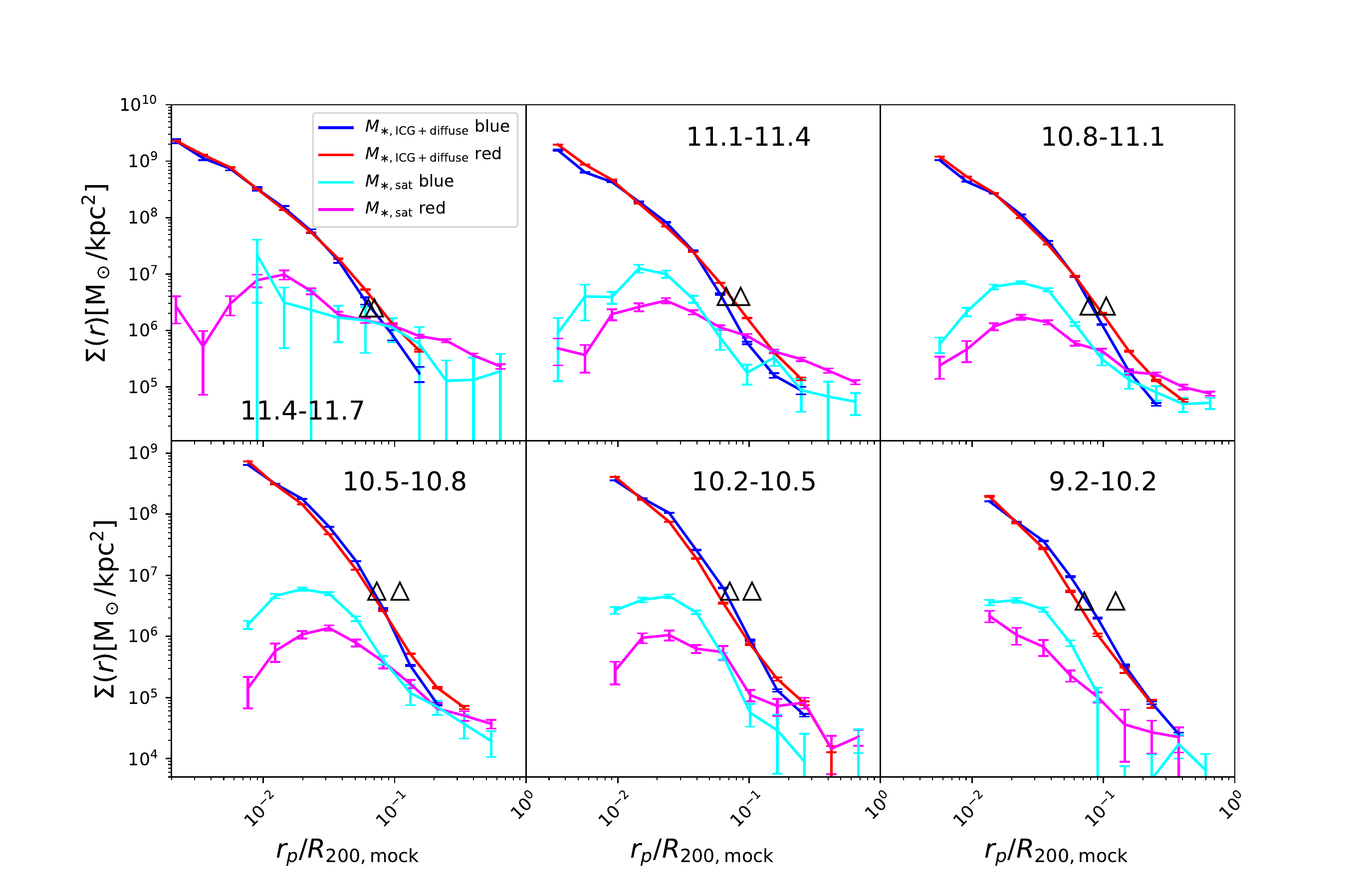}%
\caption{Projected stellar mass density profiles of ICGs $+$ their stellar halos (red and blue) and of satellite
galaxies (magenta and cyan), around ICGs with different stellar mass (see the text in each panel). In a given
panel, red/magenta and blue/cyan curves show the results around red and blue ICGs, respectively. We only include
those satellites with log stellar mass greater than $\log_{10}M_{\ast,\mathrm{ICG}}-3$, i.e., within three orders
of difference with respect to the central ICGs. Note the satellite profiles can be significantly affected by deblending
mistakes within $0.1R_{200,\mathrm{mock}}$ (see Appendix~\ref{app:deblending}). The inner and outer black triangles 
mark twice the Petrosian radius of red and blue ICGs. Blue ICGs are dominated by less concentrated late-type galaxies, 
and thus have larger Petrosian radius, but their outer stellar halos are less extended. Errorbars are based on the
1-$\sigma$ scatters of 100 boot-strap subsamples.
}
\label{fig:profmasscolor}
\end{figure*}

\begin{figure}
\includegraphics[width=0.49\textwidth]{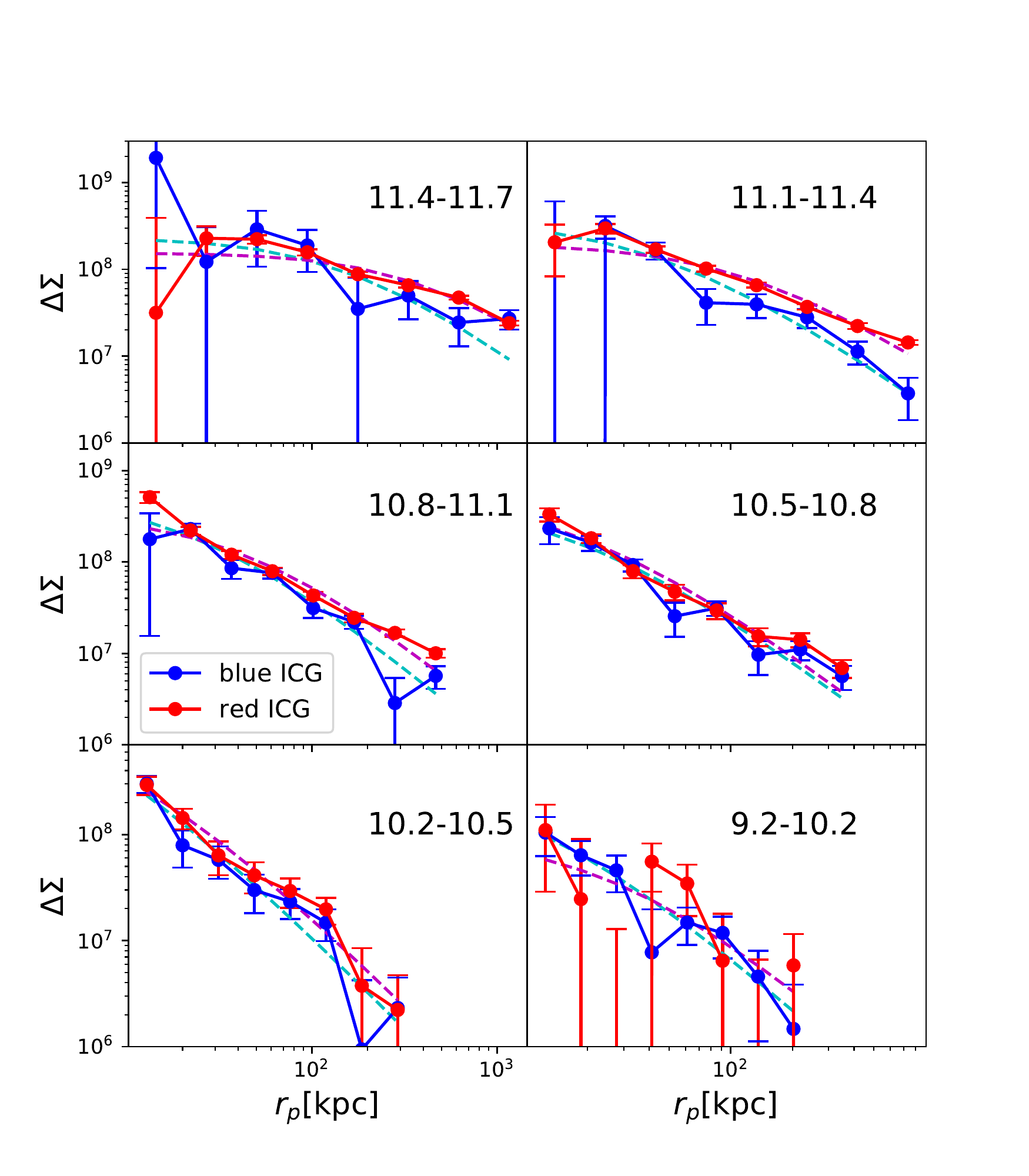}
\caption{Lensing signals around red and blue ICGs (see the legend) in a few different log stellar mass bins
(see the text in each panel). Errorbars are based on the 1-$\sigma$ scatters of 100 boot-strap subsamples. 
Dashed curves with magenta and cyan colors are the best-fitting NFW model profiles. 
}
\label{fig:profDM}
\end{figure}

In Figure~\ref{fig:profmasscolor}, the profiles centered on ICGs with different stellar mass are presented in separate
panels. In each panel, ICGs are further split into red and blue populations based on a stellar mass dependent color
division of $^{0.1}(g-r)<0.065\log_{10}M_\ast/\msun+0.1$. The profiles of satellite galaxies behave very differently
around red and blue ICGs. Beyond $0.1R_{200,\mathrm{mock}}$, where the profiles are not significantly affected by 
deblending issues, the amplitudes are higher around red ICGs with $\log_{10}M_\ast/\msun>10.8$. At $\log_{10}M_\ast/\msun<10.8$, 
we fail to see significant differences given the noisy measurements.

In an early study, \cite{2012MNRAS.424.2574W} reported that the satellite abundance around red isolated galaxies
is higher than those around blue ones with the same stellar mass, which reflects the fact that red central galaxies
are hosted by more massive dark matter halos than blue centrals with the same stellar mass. This has been directly
confirmed by weak lensing measurements from SDSS \citep{2016MNRAS.457.3200M}, and it was found that the difference
in satellite abundance and host halo mass between red and blue ICGs both peak at the stellar mass of about
$\log_{10}M_\ast/\msun\sim11.1$.

With the high quality weak gravitational lensing shear measurements based on the deep and high resolution HSC imaging
products, we can measure the lensing signals as a function of the projected radius, $r_p$, despite the fact that the
footprint of HSC is much smaller than SDSS. The lensing signals are presented in Figure~\ref{fig:profDM}. In the top
right and middle left panels ($10.8<\log_{10}M_\ast/\msun<11.4$), the lensing signals around red ICGs are clearly
higher in amplitudes. For the most massive stellar mass bin ($11.4<\log_{10}M_\ast/\msun<11.7$), the middle right
panel ($10.5<\log_{10}M_\ast/\msun<10.8$) and the bottom left panel ($10.2<\log_{10}M_\ast/\msun<10.5$), there are
also indications that the blue curve is below the red one, but the difference is not significant compared with the 
errorbars. Besides, there are no obvious differences between the red and blue curves in the two bottom panels. Note 
we do not have enough numbers of blue/red ICGs at the most/least massive ends. Nevertheless, the trends revealed by 
lensing signals are in good agreement with the results based on satellites in Figure~\ref{fig:profmasscolor}, i.e., 
over the stellar mass range of $10.8<\log_{10}M_\ast/\msun<11.4$, we clearly detect that red ICGs tend to have more 
satellites and are hosted by more massive dark matter halos than blue ICGs in the same stellar mass bin.

The profiles of ICGs $+$ their stellar halos in Figure~\ref{fig:profmasscolor} behave differently for red and blue ICGs
as well. On small scales, red ICGs tend to have slightly more concentrated profiles than blue ICGs, except for the 
most massive bin. Such a difference is more obvious for smaller ICGs in the three bottom panels, i.e., the profiles 
of blue ICGs tend to have higher amplitudes than those of red ICGs at $0.02R_{200}<r_p<0.09R_{200}$. The top middle and 
right panels also show similar but weaker trends. Note in the most massive bin, there is not enough number of blue ICGs, 
and our conversion from surface brightness profiles to projected stellar mass density profiles might have weakened such 
a difference.

On larger scales, red ICGs tend to have more extended stellar halos than blue ones. This is in good agreement with previous
studies based on both real observations \citep[e.g.][]{2014MNRAS.443.1433D} and hydrodynamical simulations \citep[e.g.][]{
2014MNRAS.444..237P,2016MNRAS.458.2371R}. The outer stellar halo is believed to be dominated by stripped stars from 
satellites, and thus the difference indicates that red passive galaxies accrete more stars than blue star-forming galaxies.

The facts that red centrals have more satellites, more extended stellar halos and are hosted by more massive dark matter
halos support the standard theory of structure and galaxy formation, which we will discuss in Section~\ref{sec:conclusion}.
Despite the difference in $M_{200}$ (hence $R_{200}$) for red and blue ICGs at fixed stellar mass, as revealed by weak
lensing signals, we scale the projected distance $r_p$ to red and blue ICGs using the same $R_{200,\mathrm{mock}}$ in 
Table~\ref{tbl:icg}, without distinguishing by color.

We note in the end that the profiles of satellites around blue ICGs tend to be more concentrated than those around 
red ICGs for the few inner most data points, which might indicate the stronger tidal disruption around red ICGs in 
such inner regions. However, as shown in Appendix~\ref{app:deblending}, the satellite counts may have been significantly 
affected by deblending issues within $0.1R_{200,\mathrm{mock}}$. In fact, \cite{2021MNRAS.500.3776W} have reported that 
the rich substructures of blue late-type ICGs, such as spiral arms and star-forming regions, can have high possibilities 
to be mistakenly deblended as companion sources, which result in faked increase in the inner number density profiles of 
satellites. Proper investigations of the inner satellite profiles require very careful corrections for such deblending 
issues.

By fitting the NFW profiles to the lensing signals in Figure~\ref{fig:profDM}, we are able to recover the best-fitting
virial mass, $M_{200}$, for the host halos of ICGs with different stellar mass, which can be used to directly investigate 
the relations between host halo mass and the total stellar mass in satellites, in ICGs and their extended stellar halos. 
In the next section, we move on to investigate these scaling relations for red and blue ICGs separately.

\subsection{Scaling Relations}
\label{sec:scaling}

\begin{figure}
\includegraphics[width=0.49\textwidth]{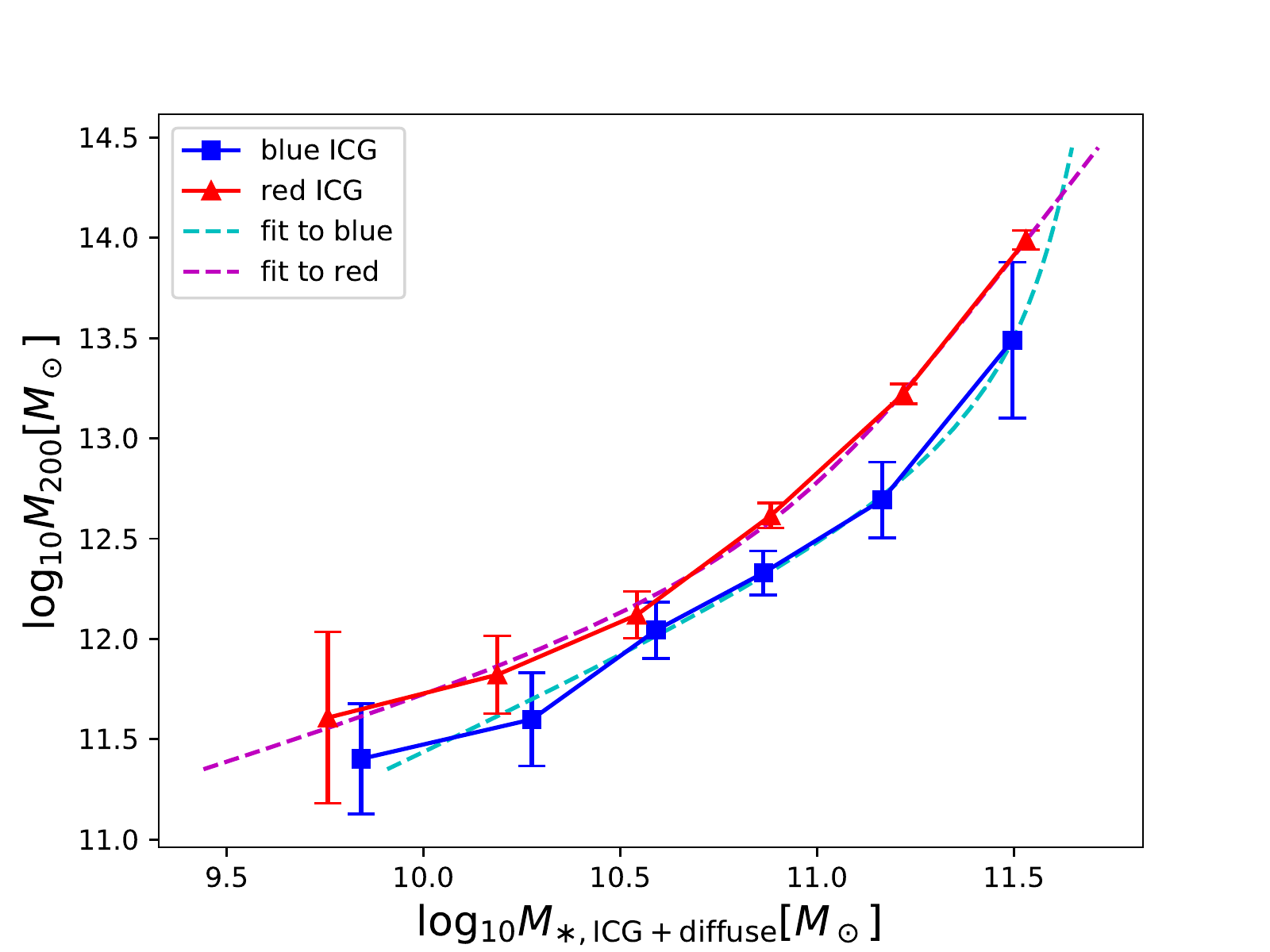}
\caption{Virial mass for the host halos of ICGs, $M_{200}$, measured from the stacked weak lensing signals in Figure~\ref{fig:profDM},
versus the integrated stellar mass for ICGs $+$ their stellar halos. Magenta and cyan dashed curves are best-fitting double 
power-law functions to the measurements for red and blue ICGs, with the best-fitting parameters provided in Table~\ref{tbl:scaling}. 
If fitting single power-law models to the four most massive data points, the best-fitting results are $M_{200}\propto
M_{\ast,\mathrm{ICG+diffuse}}^{2.064 \pm 0.080}$ and $M_{200}\propto M_{\ast,\mathrm{ICG+diffuse}}^{2.094 \pm 0.512}$ 
for red and blue ICGs, respectively.
}
\label{fig:halovsstellar}
\end{figure}

\begin{figure}
\includegraphics[width=0.49\textwidth]{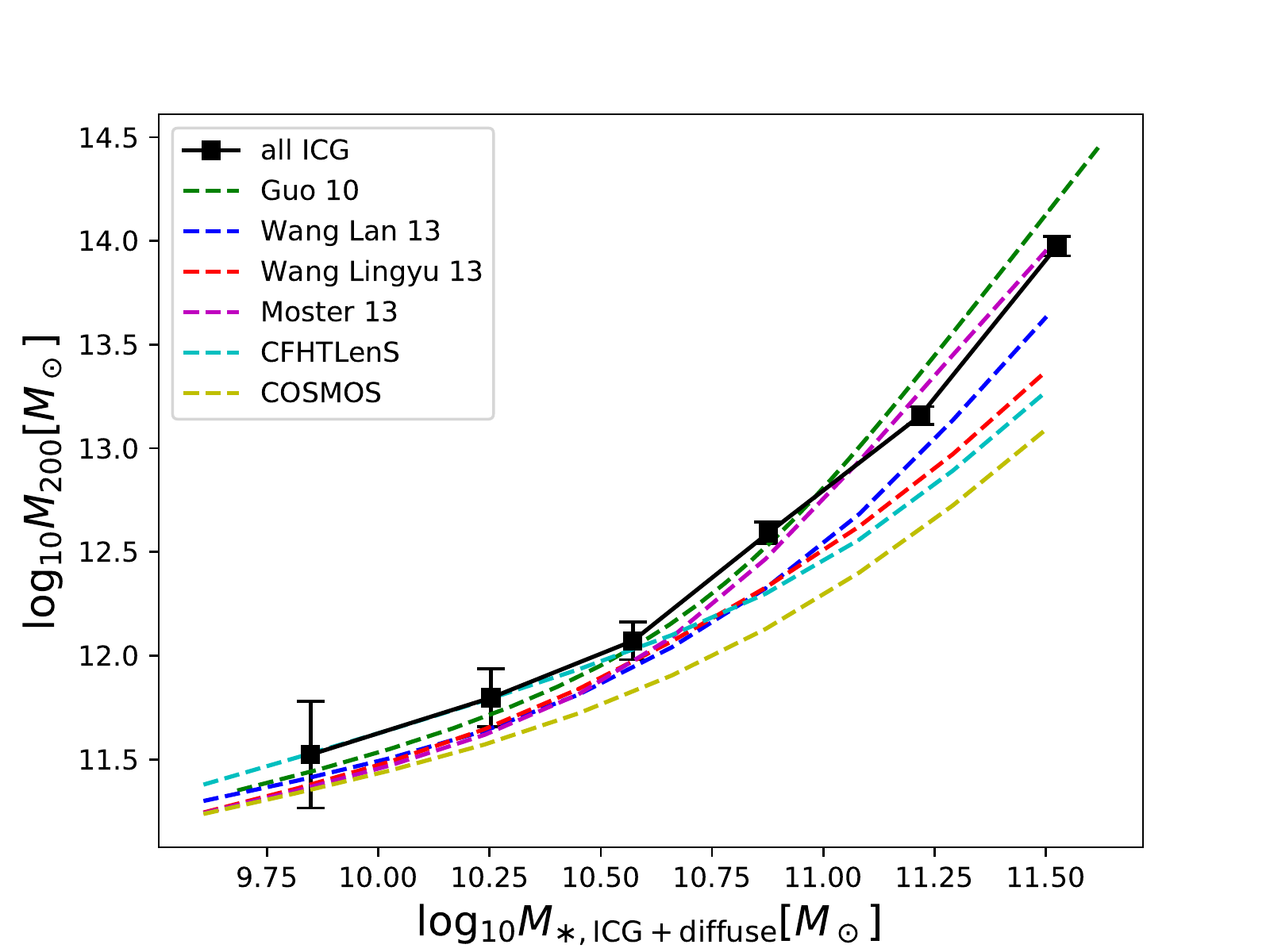}
\caption{Similar to Figure~\ref{fig:halovsstellar}, but black squares with errorbars show the relation of $M_{200}$ 
versus $M_{\ast,\mathrm{ICG+diffuse}}$ for all ICGs. Dashed curves with different colors are stellar mass versus halo 
mass relations in a few previous studies \citep{2010MNRAS.404.1111G,2012ApJ...744..159L,2013MNRAS.431..648W,
2013MNRAS.428.3121M,2013MNRAS.431..600W,2015MNRAS.447..298H}.
}
\label{fig:halovsstellarall}
\end{figure}

\begin{figure}
\includegraphics[width=0.49\textwidth]{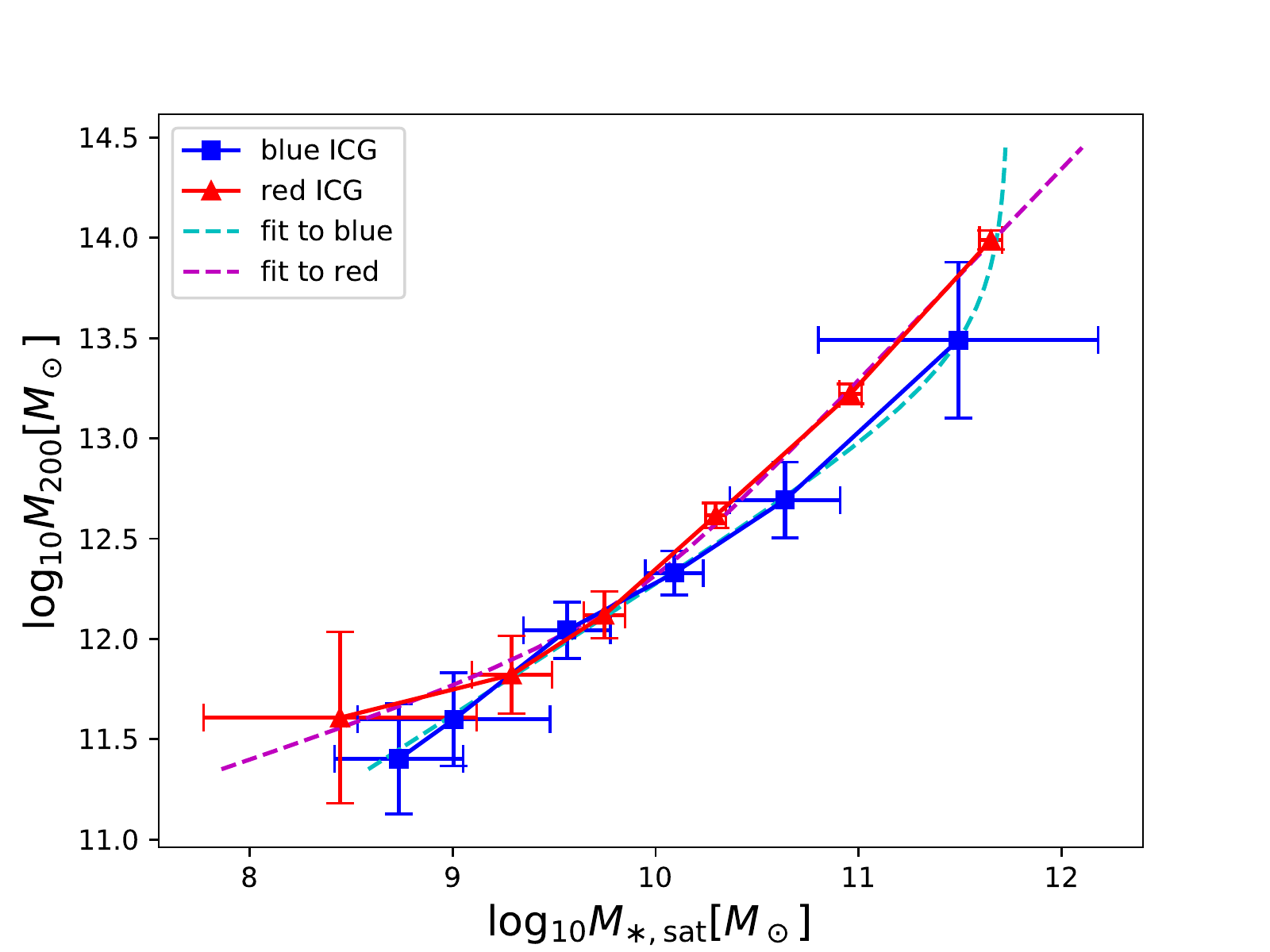}
\caption{Similar to Figure~\ref{fig:halovsstellar}, but $M_{200}$ is shown as a function of the stellar mass in satellites
with $\log_{10}M_{\ast,\mathrm{sat}}>\log_{10}M_{\ast,\mathrm{ICG}}-3$ and with projected distance to the central ICGs in between
0.1$R_{200,\mathrm{mock}}$ and $R_{200,\mathrm{mock}}$. ICGs are split into red and blue populations, as indicated by the legend. 
Magenta and cyan dashed curves are best-fitting double power-law functions to the measurements for red and blue ICGs, with 
the best-fitting parameters provided in Table~\ref{tbl:scaling}. If fitting single power-law models to the four most massive data 
points, the best-fitting results are $M_{200}\propto M_{\ast,\mathrm{sat}}^{0.976\pm 0.036}$ and $M_{200}\propto M_{\ast,\mathrm{sat}}
^{1.017 \pm 0.184}$ for red and blue ICGs, respectively.
}
\label{fig:halovssat}
\end{figure}

\begin{figure}
\includegraphics[width=0.49\textwidth]{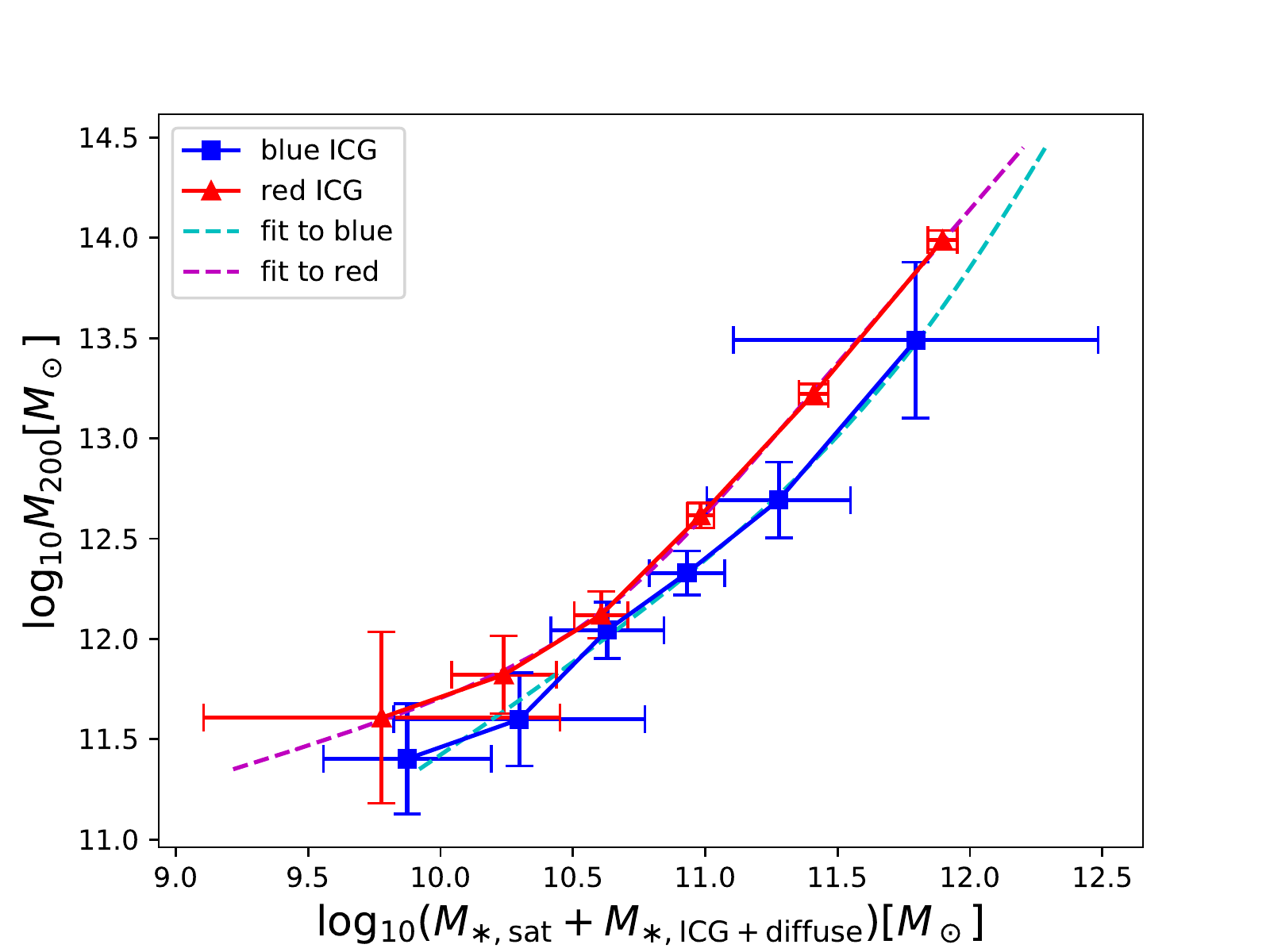}
\caption{$M_{200}$ as a function of the total stellar mass in satellites and ICGs $+$ their stellar halos. Similar
to Figure~\ref{fig:halovssat}, we only include satellites with $\log_{10}M_{\ast,\mathrm{sat}}>\log_{10}M_{\ast,
\mathrm{ICG}}-3$ and with projected distance to the central ICGs in between 0.1$R_{200,\mathrm{mock}}$ and $R_{200,
\mathrm{mock}}$. ICGs are split into red and blue populations, as indicated by the legend. Magenta and cyan 
dashed curves are best-fitting double power-law functions to the measurements for red and blue ICGs, with the 
best-fitting parameters provided in Table~\ref{tbl:scaling}. If fitting single power-law models to the four most 
massive data points, the best-fitting results are $M_{200}\propto (M_{\ast,\mathrm{sat}}+M_{\ast,\mathrm{ICG+diffuse}})
^{1.449\pm 0.041}$ and $M_{200}\propto (M_{\ast,\mathrm{sat}} +M_{\ast,\mathrm{ICG+diffuse}})^{1.676\pm 0.297}$ 
for red and blue ICGs, respectively.
}
\label{fig:halovstotal}
\end{figure}

\begin{figure*}
\includegraphics[width=0.49\textwidth]{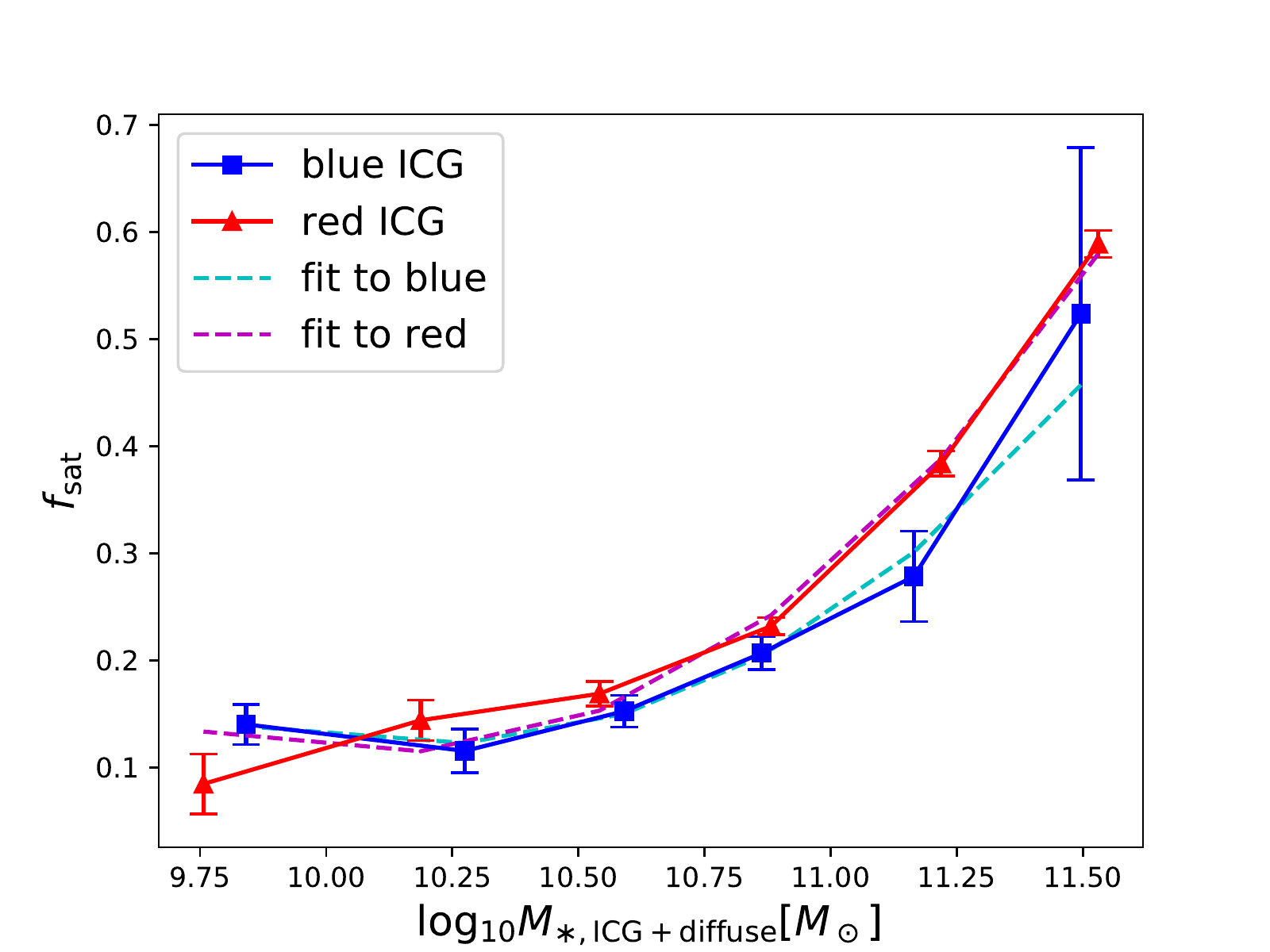}%
\includegraphics[width=0.49\textwidth]{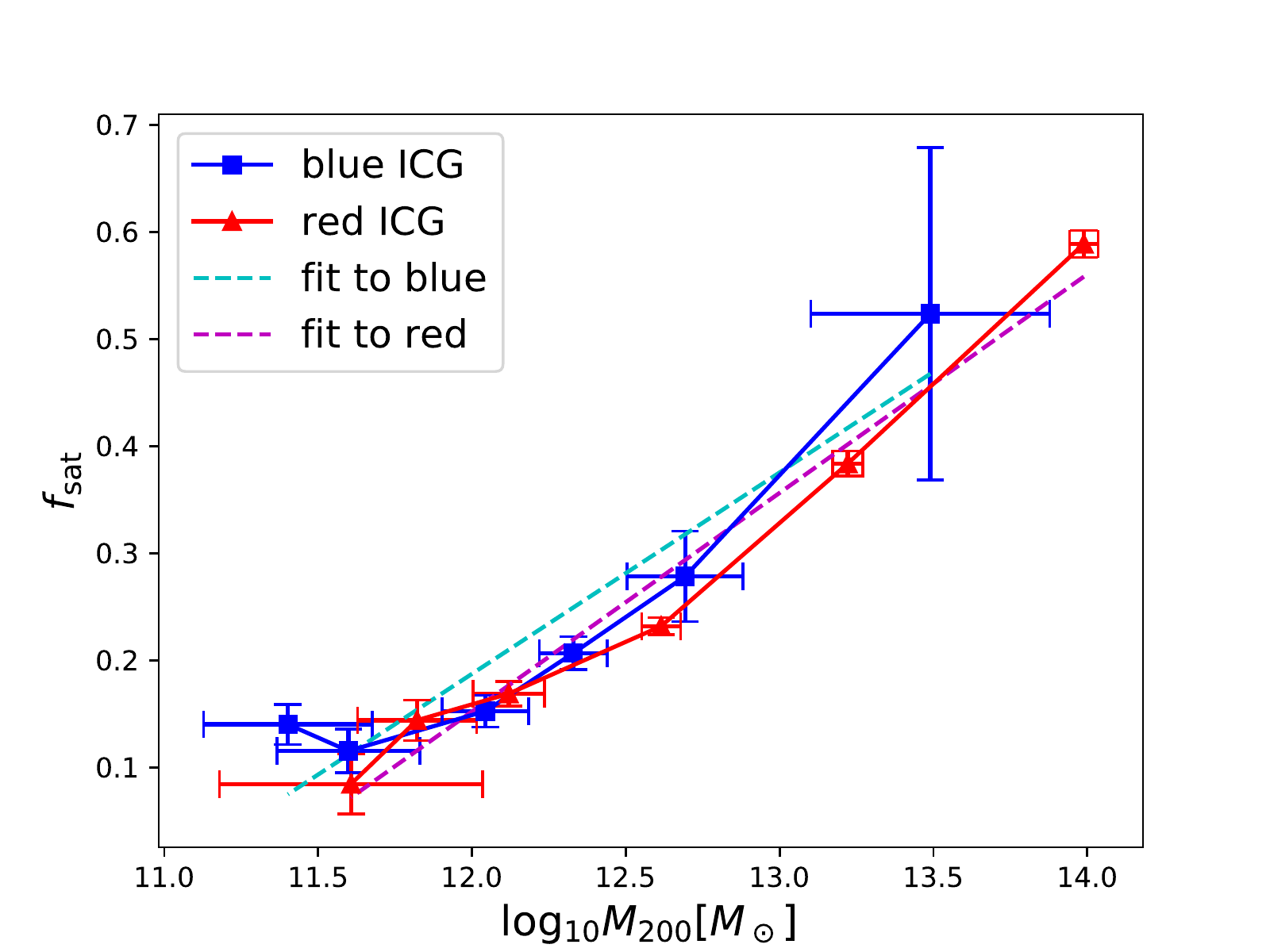}%
\caption{The fraction of stellar mass in satellites versus the total stellar mass in satellites and ICGs $+$ their
stellar halos, reported as a function of the stellar mass in ICGs and their stellar halos, $M_{\ast,\mathrm{ICG+diffuse}}$,
(left), and as a function of $M_{200}$ (right). Red and blue symbols and curves are for red and blue ICGs, respectively.
Magenta and cyan curves are best-fitting cubic polynomial models (left) and power-law models (right) for red and 
blue ICGs.
}
\label{fig:satfrac}
\end{figure*}

The weak lensing mass profiles in Figure~\ref{fig:profDM} are measured and fitted out to $2 R_{200,\mathrm{mock}}$,
which is close to the halo depletion radius~\citep{FH21} that marks the separation between the one and two halo terms 
in halo model~\citep[e.g.,][]{HayashiWhite,Garcia20}. For each measured profile we fit a single projected NFW profile
\citep{2000ApJ...534...34W} with the virial mass, $M_{200}$, and concentration, $c_{200}$, as free parameters. The 
fitting is done by minimizing the value of $\chi^2$ using the software \textsc{iminuit}, which is a python interface 
of the \textsc{minuit} function minimizer \citep{1975CoPhC..10..343J}. We have also tried the \textsc{emcee} software 
\citep{2013PASP..125..306F}. The best-fitting NFW profiles and the associated errors by using \textsc{iminuit} and 
\text{emcee} agree well with each other. 

In principle, we can directly use $R_{200}$ inferred from weak lensing signals, which are not exactly the same as
$R_{200,\mathrm{mock}}$ in Table~\ref{tbl:icg}. In fact, $R_{200}$ inferred from weak lensing signals are larger 
than $R_{200,\mathrm{mock}}$ by about 24\%, 9\%, 12\%, 1.6\%, 2.2\% and $<$1\% from the most to least massive bins. 
The discrepancy is larger for more massive bins, which is perhaps related to the large scatter in $M_{200}$ at fixed 
stellar mass for massive halos. However, given the measurement uncertainties, we stick to use $R_{200,\mathrm{mock}}$ 
in Table~\ref{tbl:icg}. It affects how the results are presented in Figures~\ref{fig:profmassall} and \ref{fig:profmasscolor}, 
but does not change the conclusions. As we have checked, the trends in Figures~\ref{fig:profmassall} slightly differ, 
but the conclusion still holds if using $R_{200}$ estimated from weak lensing signals. The other conclusions which 
are going to be presented in this subsection are not affected at all.

Figure~\ref{fig:halovsstellar} shows the best-fitting virial mass of the host dark matter halo, $M_{200}$, versus the
stellar mass of ICGs $+$ their stellar halos, $M_{\ast,\mathrm{ICG+diffuse}}$, which is integrated over the profiles
in Figure~\ref{fig:profmasscolor}. Note here we choose to use $M_{\ast,\mathrm{ICG+diffuse}}$, instead of the original
stellar mass of ICGs measured from the shallower SDSS photometry, which we denote as $M_{\ast,\mathrm{ICG}}$. We only
show the vertical errorbars for the best-fitting $M_{200}$ from weak lensing signals, as the errorbars in $M_{\ast,
\mathrm{ICG+diffuse}}$ are very small for ICGs in a given stellar mass bin.

It seems $M_{200}$ of red ICGs is on average larger than that of blue ICGs in the same stellar mass bin, but only the
second and third massive data points show more than 1-$\sigma$ significance, which is in very good agreement with the
trends shown in Figures~\ref{fig:profmasscolor} and \ref{fig:profDM}. At $\log_{10}M_{\ast,\mathrm{ICG+diffuse}}/\msun
>10.5$, the power-law index is very close to 2 (or $M_{\ast,\mathrm{ICG+diffuse}} \propto M_{200}^{1/2}$). Indeed, for
the four most massive data points of red ICGs, where the errorbars are small enough, the best-fitting power-law index
is $2.064 \pm 0.080$. At $\log_{10}M_{\ast,\mathrm{ICG+diffuse}}/\msun<10.5$, on the contrary, the amount of change
in host halo mass becomes slower, despite the significant decrease in stellar mass. The general trends are consistent 
with stellar mass versus halo mass relations measured in previous studies \citep[e.g.][]{2010MNRAS.404.1111G,2010MNRAS.402.1796W,2012ApJ...752...41Y,2012ApJ...744..159L,2013MNRAS.428.3121M,
2013MNRAS.431..648W,2013MNRAS.431..600W,2015MNRAS.447..298H}.

Following the functional form adopted in \cite{2010MNRAS.402.1796W}, we model the relation of $M_{200}$ versus 
$M_{\ast,\mathrm{ICG+diffuse}}$ for red and blue ICGs separately as

\begin{equation}
    M_{\ast,\mathrm{ICG+diffuse}}=\frac{2k}{(\frac{M_{200}}{M_0})^{-\alpha}+(\frac{M_{200}}{M_0})^{-\beta}}.
    \label{eqn:doublepower}
\end{equation}

In order to properly consider the errors in $M_{200}$ for the fitting, Equation~\ref{eqn:doublepower} is in fact 
tabulated into values of $M_{200}$ versus $M_{\ast,\mathrm{ICG+diffuse}}$, for a given set of parameters, before 
fitting to the data points. The best-fitting relations are shown as magenta and cyan dashed curves in 
Figure~\ref{fig:halovsstellar}, for red and blue ICGs respectively. The best-fitting parameters are provided 
in the first and second rows of Table~\ref{tbl:scaling}.

Our results in Figure~\ref{fig:halovsstellar} are consistent with the commonly
recognized galaxy formation models. For massive galaxies, a change in stellar mass leads to a more rapid change
in host halo mass. In standard galaxy formation theory, AGN feedback is often adopted to prohibit the star-forming
activities in massive galaxies, making the growth in stellar mass less efficient, which might explain why with
the change in host halo mass, the change in stellar mass is smaller. For galaxies smaller than $\log_{10}M_{\ast,
\mathrm{ICG+diffuse}}/\msun\sim10.5$, on the other hand, the slow change in host halo mass indicates that the star
formation activities in low-mass halos are significantly more inefficient and stochastic, which depends very weakly
on host halo mass.

\begin{table*}
\caption{The best-fitting parameters for the scaling relations presented in Figures~\ref{fig:halovsstellar},
\ref{fig:halovssat} and \ref{fig:halovstotal}, and for red and blue ICGs respectively. The model functional form is 
$X=\frac{2k}{(\frac{M_{200}}{M_0})^{-\alpha}+(\frac{M_{200}}{M_0})^{-\beta}}$, where $X$ represents $M_{\ast,\mathrm{ICG+diffuse}}$, 
$M_{\ast,\mathrm{sat}}$ or $M_{\ast,\mathrm{ICG+diffuse}}+M_{\ast,\mathrm{sat}}$. Note the data points around red ICGs 
tend to have higher amplitudes in corresponding figures, but the best-fitting amplitudes here are lower. This is due to the 
positive correlation between $\log_{10}k$ and $\log_{10}M_0$, and the negative correlation between $\log_{10}k$ and $\beta$.} 
\begin{center}
\begin{tabular}{lcccc}\hline\hline
relation  & $\log_{10}M_0$ & $\alpha$ & $\beta$ & $\log_{10}k$\\ \hline
$M_{200}$ vs $M_{\ast,\mathrm{ICG+diffuse}}$ (red ICG)  & 12.14$\pm$0.50 &  1.69$\pm$1.13  & 0.39$\pm$0.13  &  10.51$\pm$0.43  \\
$M_{200}$ vs $M_{\ast,\mathrm{ICG+diffuse}}$ (blue ICG) & 12.80$\pm$2.58  & 1.10$\pm$0.86  & 0.08$\pm$3.81  & 11.21$\pm$1.31  \\
$M_{200}$ vs $M_{\ast,\mathrm{sat}}$ (red ICG)  &  11.97$\pm$0.86 &  3.00$\pm$5.47  & 0.95$\pm$0.26  &  9.45$\pm$1.25 \\
$M_{200}$ vs $M_{\ast,\mathrm{sat}}$ (blue ICG) &  13.39$\pm$1.06  & 1.54$\pm$0.66  & 0.01$\pm$4.86  & 11.42$\pm$0.83 \\
$M_{200}$ vs $M_{\ast,\mathrm{ICG+diffuse}}+M_{\ast,\mathrm{sat}}$ (red ICG)  & 11.80$\pm$0.48  & 2.67$\pm$1.70  & 0.65$\pm$0.12  & 10.18$\pm$0.50  \\
$M_{200}$ vs $M_{\ast,\mathrm{ICG+diffuse}}+M_{\ast,\mathrm{sat}}$ (blue ICG)   & 12.74$\pm$1.94 &  1.18$\pm$0.72   &  0.42$\pm$1.06  &  11.29$\pm$1.72\\
\hline
\label{tbl:scaling}
\end{tabular}
\end{center}
\end{table*}

\begin{table*}
\caption{The best-fitting parameters for $f_\mathrm{sat}$ versus $M_{\ast,\mathrm{ICG+diffuse}}$ and $f_\mathrm{sat}$ versus
or $M_{200}$ in Figure~\ref{fig:satfrac}, and for red and blue ICGs respectively. The model functional form is demonstrated 
by Equations~\ref{eqn:cubic} and \ref{eqn:singlepower}.} 
\begin{center}
\begin{tabular}{lcccc}\hline\hline
relation  & $a$ & $b$ & $c$ & $d$\\ \hline
$f_\mathrm{sat}$ vs $M_{\ast,\mathrm{ICG+diffuse}}$ (red ICG)  & 28.4658$\pm$0.0876  & $-$67.6943$\pm$0.1190  & 49.6482$\pm$0.1304  & $-$10.3045$\pm$0.1135  \\
$f_\mathrm{sat}$ vs $M_{\ast,\mathrm{ICG+diffuse}}$ (blue ICG) & 26.4056$\pm$0.1666 &  $-$64.7616$\pm$0.2164   &  49.7637$\pm$0.2283  &  $-$11.2813$\pm$0.1928  \\
$f_\mathrm{sat}$ vs $M_{200}$ (red ICG)  & 0.204$\pm$0.016  & $-$2.296$\pm$0.205  & --  & --  \\
$f_\mathrm{sat}$ vs $M_{200}$ (blue ICG)   & 0.188$\pm$0.033 &  $-$2.068$\pm$0.403   &  --  &  --  \\
\hline
\label{tbl:scaling2}
\end{tabular}
\end{center}
\end{table*}

Figure~\ref{fig:halovsstellarall} shows a more detailed comparison of the stellar mass versus halo mass relation 
for all ICGs against the measurements in a few previous studies, which are either based on abundance matching and HOD 
(Halo Occupation Distribution) modelling \citep[][]{2010MNRAS.404.1111G,2010MNRAS.402.1796W,2010ApJ...710..903M,
2013MNRAS.428.3121M,2013MNRAS.431..648W,2013MNRAS.431..600W} or based on modelling lensing signals of other surveys 
\citep{2012ApJ...744..159L,2015MNRAS.447..298H}. Note all the HOD based studies are measuring the stellar mass for 
halos at a given mass, whereas we are measuring the halo mass for ICGs in a given stellar mass bin through stacked 
lensing. Thus we make the following conversion~\citep{2015MNRAS.446.1356H}

\begin{equation}
\log_{10}M_\mathrm{halo}(M_\ast)=\int \log_{10} M_\mathrm{halo} \ud P(M_\mathrm{halo}|M_\ast),
\label{eqn:conversion0}
\end{equation}
where $\ud P(M_\mathrm{halo}|M_\ast)$ is defined as
\begin{equation}
    \ud P(M_\mathrm{halo}|M_\ast)=\frac{\ud P(M_\ast|M_\mathrm{halo})\phi(M_\mathrm{halo}) \ud M_\mathrm{halo}}{\int \ud P(M_\ast|M_\mathrm{halo})\phi(M_\mathrm{halo}) \ud M_\mathrm{halo}}.
    \label{eqn:conversion}
\end{equation}

$\phi(M_\mathrm{halo})$ in Equation~\ref{eqn:conversion} is the halo mass function. The distributions/scatters 
of stellar mass at fixed halo mass $P(M_\ast|M_\mathrm{halo})$ are taken from the original studies. Our measurement 
appears to be closest to \cite{2010MNRAS.404.1111G} and \cite{2013MNRAS.428.3121M} models, but the other
models tend to show lower amplitudes than the black squares. The dashed color curves show large variations from each
other, which are mostly due to the different amount of HOD scatter assumed in each model. If assuming the same amount 
of scatter of 0.2~dex, the discrepancy would become smaller \citep{2015MNRAS.446.1356H} among these models, and most 
of their amplitudes are still lower than that of the black squares at $\log_{10}M_{\ast,\mathrm{ICG+diffuse}}>10.6$. 

The difference between our measurements and many of the previous studies can NOT be explained by the fact that 
we adopted $M_{\ast,\mathrm{ICG+diffuse}}$, while these previous studies did not carefully consider the stellar mass 
contained in the outer stellar halo. Their stellar mass is expected to be lower than our $M_{\ast,\mathrm{ICG+diffuse}}$ 
at the massive end (see Appendix~\ref{app:stellarmasstest} for details), so the tension would only become 
larger if considering such a difference in stellar mass. 

This could be related to the sample selection, as the ICGs used in this study is a flux limited sample, which 
will introduce slightly over-estimated halo mass than using a volume limited sample \citep{2015MNRAS.446.1356H}.
However, after including volume corrections to our sample of ICGs, the amount of change is at most about 
$-$0.04~dex in $M_{200}$, which is far from enough to explain the difference. The other more important aspect is 
the scatter/dispersion in halo mass at fixed stellar mass in our analysis. \cite{2005MNRAS.362.1451M} reported 
that the best recovered halo mass through stacked lensing signals lies in between the actual mean and median 
values. Proper comparisons between our halo mass measurements and previous studies require careful calibrations 
of the bias introduced by such mass dispersions. A 0.5/0.7~dex of dispersion in halo mass would cause a bias from 
the median as large as $\sim-$0.17/$-$0.32~dex\citep{2015MNRAS.446.1356H}. Checked against the mock galaxy catalog 
of \cite{2011MNRAS.413..101G}, the typical scatter of halo mass in the three most massive stellar mass bins in 
our analysis ranges from 0.33 to 0.41~dex. Thus the halo mass dispersion may help to explain part of the difference 
at the massive end. The remaining source of uncertainties is at least partly contributed by the amount of 
assumed HOD scatter in Equation~\ref{eqn:conversion0}.

Figure~\ref{fig:halovssat} shows $M_{200}$ versus the integrated stellar mass in satellites, $M_{\ast,\mathrm{sat}}$.
$M_{\ast,\mathrm{sat}}$ is integrated over the radial range of $0.1R_{200,\mathrm{mock}}<r_p<R_{200,\mathrm{mock}}$ in
Figure~\ref{fig:profmasscolor}. Here we adopt an inner radius cut of $0.1R_{200,\mathrm{mock}}$ to avoid the region 
significantly affected by deblending mistakes, though the integrated stellar mass in satellites is expected to be 
dominated by the mass in the outer region and thus might not be sensitive to the inner profiles. Besides, only satellites 
with $\log_{10}M_\ast>\log_{10}M_{\ast,\mathrm{ICG}}-3$ are included, and thus in principle, $M_{\ast,\mathrm{sat}}$ 
is a lower limit. Despite this, as long as the faint-end slopes of satellite luminosity functions are shallower than $-2$, 
satellites smaller than $\log_{10}M_\ast\sim \log_{10}M_{\ast,\mathrm{ICG}}-3$ are unlikely to contribute a significant 
fraction to the total stellar mass in satellites. However, the readers may argue that the chosen mass limit for satellites 
depends on the stellar mass of ICGs, which might include some artificial dependence on the properties of central ICGs. 
We think the total stellar mass in satellites below the cut is small anyway, but we have repeated our analysis by choosing 
a fixed cut in stellar mass, and our conclusions in the following remain the same.

There is a very tight correlation between $M_{200}$ and $M_{\ast,\mathrm{sat}}$. Compared with the $M_{200}$ 
versus $M_{\ast,\mathrm{ICG+diffuse}}$ relation, the relation between $M_{200}$ and $M_{\ast,\mathrm{sat}}$ 
shows much weaker dependence on the color of ICGs. The best-fitting power-law index based on the four most massive 
data points of red ICGs at $\log_{10}M_{\ast,\mathrm{sat}}>10$ (or $\log_{10}M_{\ast,\mathrm{ICG+diffuse}}/\msun>10.5$) 
is $0.976\pm 0.036$, which is very close to 1, i.e., $M_{\ast,\mathrm{sat}}\propto M_{200}$. In addition, we also 
fit Equation~\ref{eqn:doublepower} to Figure~\ref{fig:halovssat}, and the best-fitting parameters are provided in 
Table~\ref{tbl:scaling}. The power-law index is nearly twice of that between $M_{\ast,\mathrm{ICG+diffuse}}$ and 
$M_{200}$. This implies that with the same amount of change in host halo mass, the change in the total stellar 
mass locked in satellites is more rapid than the change in the total stellar mass in central ICGs at 
$\log_{10}M_{\ast,\mathrm{ICG+diffuse}} /\msun>10.5$. 

Interestingly, we can see some indications that there are slightly more stellar mass locked in satellites 
around blue ICGs at $\log_{10}M_{200}/\msun>12.7$, but the errorbars are also very large and the difference 
is only marginal at $\log_{10}M_{200}/\msun\sim12.7$. This perhaps indicates the late formation time of 
blue galaxies, and hence retaining more substructures and satellites. However, considering the fact that 
the best-fitting $M_{200}$ are achieved through binning in stellar mass, while red and blue ICGs at fixed 
stellar mass can have different scatters in $M_{200}$, we do not make a very strong conclusion here.

For completeness, we show in Figure~\ref{fig:halovstotal} the halo mass, $M_{200}$, versus the total stellar
mass in satellites and central ICGs $+$ their stellar halos, $M_{\ast,\mathrm{sat}}+M_{\ast,\mathrm{ICG+diffuse}}$.
Based on the four most massive data points of red ICGs, the best-fitting power-law index is $1.449 \pm 0.041$
($M_{\ast,\mathrm{sat}}+M_{\ast,\mathrm{cen+diffuse}} \propto M_{200}^{1/1.449}$). The best-fitting parameters 
based on Equation~\ref{eqn:doublepower} are provided in Table~\ref{tbl:scaling}. The power-law index is in 
between that of $M_{\ast,\mathrm{ICG+diffuse}}$ versus $M_{200}$ and that of of $M_{\ast,\mathrm{sat}}$ versus 
$M_{200}$, and is expected to depend on the fraction of stellar mass locked in satellites.

The left plot of Figure~\ref{fig:satfrac} shows the fraction of stellar mass in satellites versus the
total stellar mass in satellites and ICGs $+$ their stellar halos, $f_\mathrm{sat}$, reported as a
function of $M_{\ast,\mathrm{ICG+diffuse}}$, while the right plot shows the same fraction as a function
of $M_{200}$.

In both plots, $f_\mathrm{sat}$ increases with the increase in $M_{\ast,\mathrm{ICG+diffuse}}$ and $M_{200}$
for the four most massive data points. $f_\mathrm{sat}$ can be as high as 50--60\% at the massive end, 
in general consistency with previous studies based on galaxy clusters \citep[e.g.]{2013ApJ...778...14G,
2021MNRAS.502.2419F}, predictions by hydro-dynamical simulations \citep[e.g.]{2010MNRAS.406..936P,
2014MNRAS.437..816C} and HOD modelling\citep[e.g.][]{2009ApJ...693..830Y}, though in these previous studies, 
there are still some tensions between observational constraints and predictions by simulations. For MW-mass 
galaxies, $f_\mathrm{sat}$ drops to slightly more than $\sim10\%$. At the low-mass end of the left plot, 
$f_\mathrm{sat}$ is nearly flat, again indicating the more stochastic star formation of low-mass galaxies 
and also the stochastic accretion of low-mass satellites by low-mass central ICGs. Interestingly, in 
contrast to the slow change of $M_{200}$ with the decrease in both $M_{\ast,\mathrm{ICG+diffuse}}$ and 
$M_{\ast,\mathrm{sat}}$, $f_\mathrm{sat}$ tends to have a more linear relation with $M_{200}$ in the right plot, 
though the best-fitting $M_{200}$ is quite noisy at the low-mass end.

In the left plot, $f_\mathrm{sat}$ is higher around red ICGs than blue ICGs at fixed stellar mass, except for the
least massive bin. The trends are self-consistent and can be explained by the fact that red galaxies sit in denser 
environments of our Universe and thus accrete more massive satellites.

Interestingly, the right plot of Figure~\ref{fig:satfrac} reveals that the fractions of stellar mass in 
satellites around red and blue ICGs are similar, but at $\log_{10}M_{200}/\msun>12.7$, it seems
$f_\mathrm{sat}$ is slightly lower for red ICGs than blue ICGs hosted by dark matter halos with the same $M_{200}$. 
This perhaps tells that the fraction of stellar mass in satellites is mainly determined by the host halo mass, 
while in agreement with Figure~\ref{fig:halovssat}, it probably also reflects the late formation of blue 
galaxies, which is a secondary factor compared with the host halo mass. However, the difference is not 
significant compared with the errorbars.

We also provide the best-fitting relations of $f_\mathrm{sat}$ versus $M_{\ast,\mathrm{ICG+diffuse}}$ or $M_{200}$.
Cubic polynomial model is adopted to fit the relation between $f_\mathrm{sat}$ and $M_{\ast,\mathrm{ICG+diffuse}}$

\begin{eqnarray}
    f_\mathrm{sat}=&&a \left(\frac{\log_{10}M_{\ast,\mathrm{ICG+diffuse}}}{10}\right)^3+b \left(\frac{\log_{10}M_{\ast,\mathrm{ICG+diffuse}}}{10}\right)^2 \nonumber \\
    &&+ c \left(\frac{\log_{10}M_{\ast,\mathrm{ICG+diffuse}}}{10}\right)+d,
    \label{eqn:cubic}
\end{eqnarray}
while single power-law model is adopted to fit the relation between $f_\mathrm{sat}$ and $M_{200}$

\begin{equation}
    f_\mathrm{sat}=a \log_{10}M_{200} +b.
    \label{eqn:singlepower}
\end{equation}

The best-fitting parameters are provided in Table~\ref{tbl:scaling2}.

\subsection{Radial dependence of the total stellar mass fraction}

\begin{figure}
\includegraphics[width=0.49\textwidth]{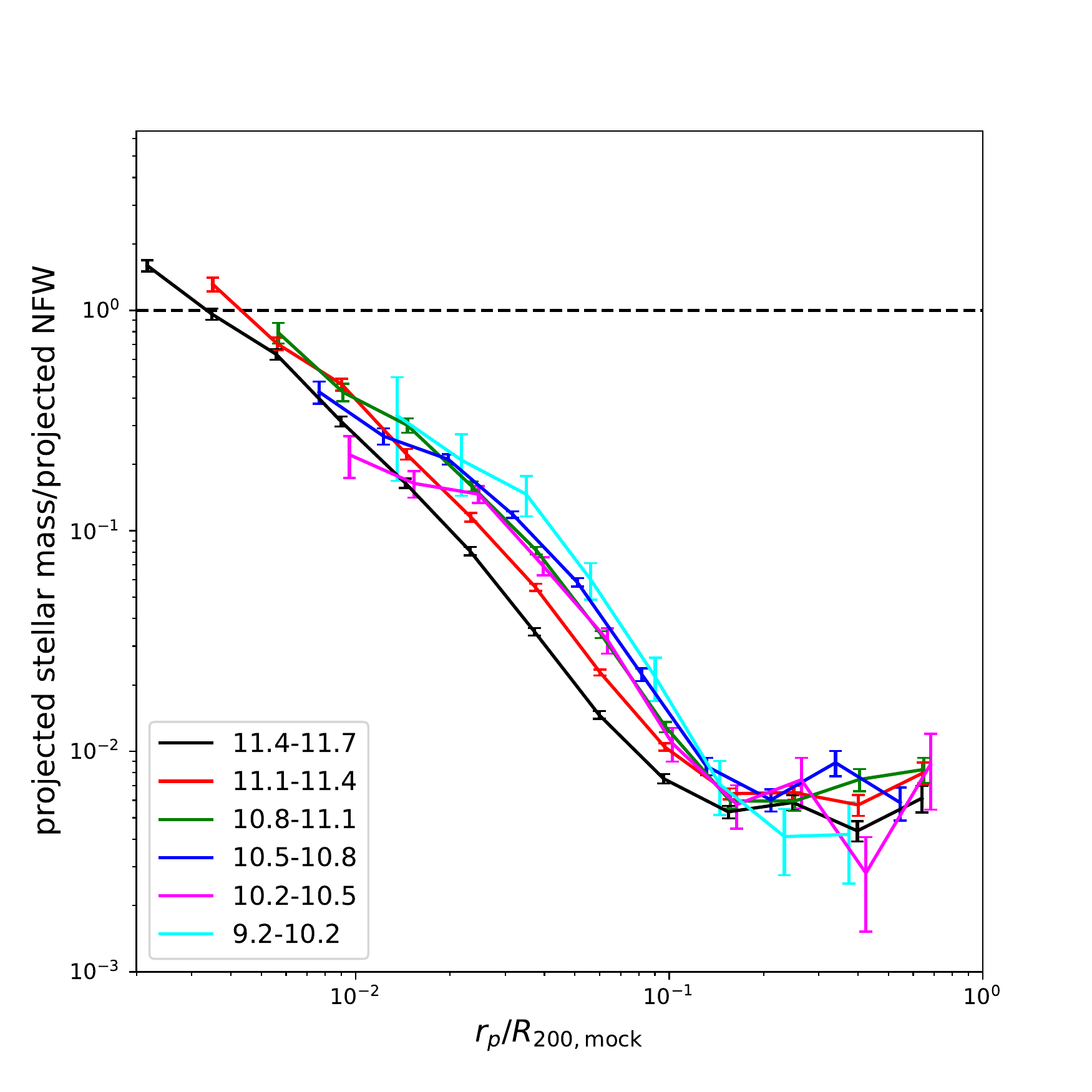}
\caption{Total stellar mass (stellar mass in satellites and ICGs $+$ their stellar halos) versus total mass
(the best-fitting NFW model profiles through weak lensing signals), as a function of  $r_p/R_{200,\mathrm{mock}}$ 
and for ICGs in different stellar mass bins. The black dashed horizontal line marks the value of unity. Errorbars
include the contribution from three different parts: (i) the boot-strap errors of satellite profiles; (ii)
the boot-strap errors of ICG $+$ stellar halo profiles; (iii) uncertainties in the best-fitting NFW model.
(i), (ii) and (iii) are all propagated to the final errors.
}
\label{fig:fracall}
\end{figure}

\begin{figure*}
\begin{center}
\includegraphics[width=0.85\textwidth]{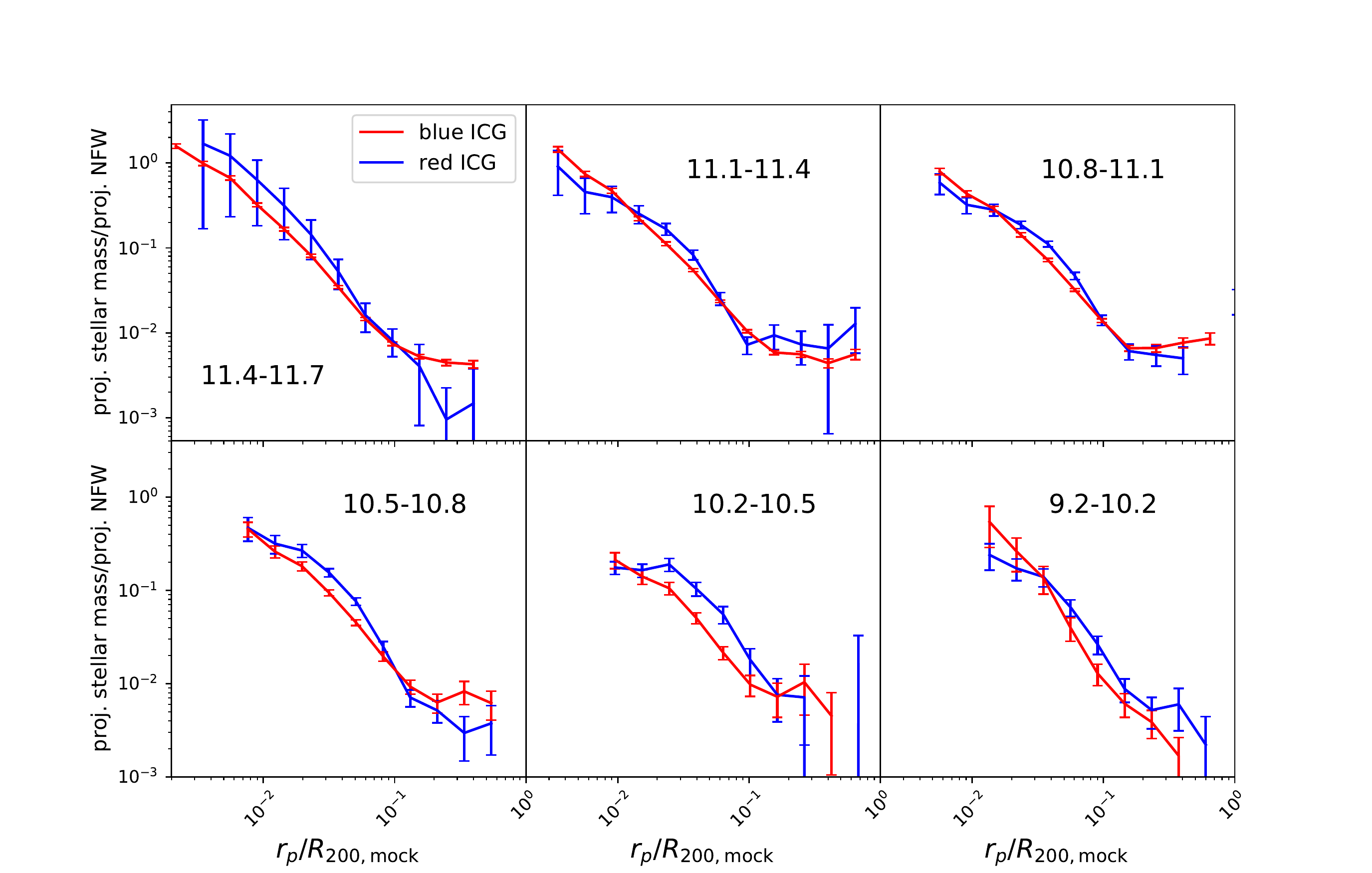}
\caption{Similar to Figure~\ref{fig:fracall}, but shows the radial distribution for the fraction of total 
stellar mass versus total mass for red and blue ICGs separately. ICGs in different stellar mass bins are 
presented in different panels, as indicated by the text in each panel. Errorbars are calculated in the same 
way as Figure~\ref{fig:fracall}.
}
\label{fig:fracredblue}
\end{center}
\end{figure*}

So far we have investigated the scaling relations of $M_{200}$ versus $M_{\ast,\mathrm{ICG+diffuse}}$, 
$M_{\ast,\mathrm{sat}}$ and versus $M_{\ast,\mathrm{ICG+diffuse}}+M_{\ast,\mathrm{sat}}$. These are 
based on the integrated mass. With the measured mass profiles, we are able to more closely investigate 
the total stellar mass fraction as a function of the projected distance to ICGs.

Figure~\ref{fig:fracall} and \ref{fig:fracredblue} show the projected total stellar mass density profiles
($M_{\ast,\mathrm{ICG+diffuse}}$ $+$ $M_{\ast,\mathrm{sat}}$ profiles), divided by the best-fitting projected
NFW density profiles based on weak lensing signals in Figure~\ref{fig:profDM}. Figure~\ref{fig:fracall} is
based on all ICGs, while in Figure~\ref{fig:fracredblue} we further divide ICGs into red and blue subsamples.
Note the inner satellite profiles within $0.1R_{200,\mathrm{mock}}$ are affected by deblending issues and are 
thus not reliable. However, in such inner regions, the profiles are dominated by the central ICGs, and thus 
the summation of the two is not sensitive to deblending mistakes.

In both figures, uncertainties of the satellite profiles, of the profiles of ICGs $+$ their stellar halos and
of the best-fitting NFW profiles through the lensing signals are all considered and propagated to the final
errorbars. The uncertainties of the best-fitting NFW profiles are estimated from the boundaries which enclose
68\% of the MCMC chains.

In Figure~\ref{fig:fracall}, high and low-mass ICGs tend to have different stellar mass fractions, in terms of both 
the shapes and amplitudes. There is nearly a monotonic trend that less massive ICGs tend to have lower fractions 
over $0.02R_{200,\mathrm{mock}}<r_p<0.1R_{200,\mathrm{mock}}$, except for the magenta and blue curves. This is partly 
related to the fact that massive ICGs are dominated by elliptical galaxies with de Vaucouleurs profiles and are thus 
more centrally concentrated. On the other hand, low-mass ICGs are dominated by exponential profiles and are more 
extended. Thus less massive ICGs tend to have less stellar mass in central regions but more stellar mass at intermediate 
radii. In addition, the fact that ICGs with $10.5<\log_{10}M_\ast/\msun<10.8$ seem to show higher amplitudes beyond 
0.02$R_{200,\mathrm{mock}}$ than ICGs in the other neighboring bins probably indicate that $M_\ast/M_{200}$ peaks 
for MW-mass galaxies \citep[e.g.][]{2010MNRAS.404.1111G}. Note the measurement in the least massive bin (cyan curve) 
might be too noisy.

In Figure~\ref{fig:fracredblue}, blue ICGs tend to have higher stellar mass fractions between $0.02R_{200,\mathrm{mock}}$ and
$0.1R_{200,\mathrm{mock}}$. The trend is also related to the fact that blue star-forming galaxies are dominated by exponential 
profiles and are thus more extended. Besides, this might imply the higher star formation efficiency per unit halo mass for blue 
star-forming galaxies.

In both Figure~\ref{fig:fracall} and \ref{fig:fracredblue}, the values of a few inner most data points are larger than one.
This likely reflects deviations from the NFW profile in the very inner region, where baryons dominate. The fractions keep
dropping with the increase in projected distance to the central ICG up to $\sim0.15R_{200,\mathrm{mock}}$. Interestingly, 
we can see the profiles go almost flat beyond $\sim0.15R_{200,\mathrm{mock}}$, and the place where the profiles start to go 
flat is almost the same after scaling $r_p$ by $R_{200,\mathrm{mock}}$. This reflects the transition radius from ICG dominated 
regions to satellite dominated regions.

In the satellite dominated region beyond the transition radius, the stellar mass versus total mass fractions are all below
1\%. The nearly flat profiles indicate the radial distribution of satellites tend to trace the distribution of dark matter,
in good agreement with previous studies \citep[e.g.][]{2018MNRAS.475.4020W}. In Figure~\ref{fig:fracredblue}, blue massive
and red low-mass ICGs tend to have noisy measurements, which is due to the small number of galaxies with corresponding colors
in these bins. There are some small differences between the red and blue curves in the satellite dominated region, but the
trend is not monotonic. Red ICGs tend to have lower stellar mass fractions than blue ICGs in the second and third massive
bins, but higher fractions in the most massive, fourth and fifth bins. Measurements in the least massive bin is too noisy
(see Figure~\ref{fig:profDM}). Given the large errorbars and noisy measurements, we avoid making strong conclusions.

Unfortunately, due to deblending issues, we are unable to have decent comparisons between the radial distribution of
stellar mass in diffuse stellar halos and satellites. This is because beyond $0.1R_{200,\mathrm{mock}}$, where the 
results are not significantly affected by deblending mistakes, we have at most one to two data points for the profiles 
of the outer stellar halos. The radial range where we have good measurements for both satellites and ICGs $+$ their 
stellar halos, while not sensitive to deblending mistakes, is very limited. We thus postpone such a comparison to 
future studies.

\section{Conclusions and discussions}
\label{sec:conclusion}

In this study, we measured the projected stellar mass density profiles for satellite galaxies as a function of the
projected distance to the center of isolated central galaxies (ICGs) and for ICGs themselves $+$ their extended stellar
halos. The profiles of satellites are measured by counting photometric companions from the deep Hyper Suprime-Cam (HSC)
imaging survey and with statistical fore/background subtraction. The profiles of ICGs $+$ their stellar halos are obtained
by stacking galaxy images from HSC to push beyond the noise limit of individual images.

The signals can be successfully measured around ICGs spanning a wide range in stellar mass ($9.2<\log_{10}M_\ast/\msun<11.7$) 
and out to the virial radius ($R_{200,\mathrm{mock}}$), which demonstrate the power of the deep HSC survey. Despite the fact 
that the footprint of HSC is significantly smaller than that of SDSS, we are able to achievement measurements for ICGs  
that are smaller by one order of magnitude in stellar mass than previous studies based on SDSS.

We found red ICGs tend to have less extended inner profiles within 0.1$R_{200,\mathrm{mock}}$, and more extended outer 
stellar halos. Over the radial range not affected by source deblending issues, the stellar mass density profiles of satellites 
have higher amplitudes around red ICGs than blue ICGs with the same stellar mass, in good agreement with previous
studies \citep[e.g.][]{2012MNRAS.424.2574W,2014MNRAS.442.1363W}, which have reported higher satellite abundance
around red ICGs. Weak lensing signals reveal that red ICGs are hosted by more massive dark matter halos than blue ICGs
with the same stellar mass, and such a difference peaks at about $\log_{10}M_\ast/\msun\sim11.1$, indicating the satellite
abundance and total stellar mass locked in satellites are good proxies to the host halo mass.

Under the standard picture of cosmic structure formation, red galaxies formed early and grew fast at early epochs. The
rapid star formation trigerred feedback mechanisms heating the surrounding gas. As a result, late-time star-forming 
activities are prohibited due to the lack of cold gas supply, while their host dark matter halos, outer stellar halos 
and the population of satellites keep growing through accretion\footnote{Late-time dry mergers can contribute to the 
growth of both dark matter and stellar mass, while still maintain the quiescent status of the galaxy. However, the peak 
stellar to dark matter ratio is only about 3\% \citep[e.g.][]{2010MNRAS.404.1111G}, and hence dry merger mainly contributes 
to the growth of dark matter and the contribution to the late-time growth in stellar mass through dry merger events is 
expected to be minor.}. Because red galaxies are on average found in more over-dense regions than blue galaxies, they 
probably even accrete more dark matter and more massive satellites. On the other hand, star forming galaxies are 
blue. Thus at fixed host halo mass, we expect red passive galaxies, whose star forming activities were quenched a long 
time ago, to be smaller in stellar mass. In other words, if the stellar mass is the same, red galaxies are hosted by 
more massive dark matter halos and have more satellites.

We revealed a tight correlation between the stellar mass in satellites, $M_{\ast,\mathrm{sat}}$, and the best-fitting 
host halo mass through weak lensing signals, $M_{200}$, with best-fitting power-law index very close to 1 at 
$\log_{10}M_{\ast,\mathrm{ICG+diffuse}}/\msun>10.5$, i.e., $M_{\ast,\mathrm{sat}}\propto M_{200}$. On the other hand, 
the scaling relation between the stellar mass in ICGs $+$ their stellar halos and $M_{200}$ is very different, and the 
best-fitting relation has a power-law index close to $1/2$ at $M_{\ast,\mathrm{ICG+diffuse}}>10.5$, i.e., $M_{\ast,
\mathrm{ICG+diffuse}} \propto M_{200}^{1/2}$. Thus with the same amount of increase in $M_{200}$, $M_{\ast,\mathrm{sat}}$ 
increases faster than $M_{\ast,\mathrm{ICG+diffuse}}$.

At $M_{\ast,\mathrm{ICG+diffuse}}<10.5$, with the decrease in $M_{\ast,\mathrm{ICG+diffuse}}$, the change in host 
halo mass is slow. This indicates the star-forming activities in low-mass galaxies is inefficient and stochastic, and 
the accretion of low-mass satellites by low-mass central ICGs is perhaps also more stochastic than that of more massive 
galaxies.

The difference between the scaling relations of $M_{200}$ versus $M_{\ast,\mathrm{sat}}$ and of $M_{200}$ versus
$M_{\ast,\mathrm{ICG+diffuse}}$ can be intuitively understood under the framework of the standard cosmic structure
formation theory. As have been mentioned, the host dark matter halo grows in mass and size through accretion. Smaller
halos, after being accreted, become subhalos and satellites. In principle, the total stellar mass in satellites can
be affected by many factors. For example, satellites can continue forming stars after being accreted by the host dark
matter halo, and their stellar material can be stripped after infall. However, if we make a few assumptions: i) the
majority of stars are formed before infall; ii) tidal stripping can be ignored, or the amount of stripped stellar 
mass is on average proportional to the stellar mass formed before infall; iii) at fixed host halo mass, the fraction 
of stellar mass show reasonable scatter around the mean, it is then not difficult to expect that, statistically,
the total stellar mass in satellites is approximately proportional to the total accreted dark matter by the host halo.
We also note that for low-mass galaxies, the scatter in stellar mass is huge at fixed halo mass based on previous
abundance matching studies \citep[e.g.][]{2010MNRAS.404.1111G,2010MNRAS.402.1796W}, which reflects the more stochastic
star formation in low-mass galaxies and explains why at $\log_{10}M_{\ast,\mathrm{ICG+diffuse}}/\msun<10.5$, $M_{200}$
changes little with the decrease in $M_{\ast,\mathrm{sat}}$. For central galaxies, the growth in stellar mass
is contributed by both in-situ star formation and ex-situ accretion of stars from satellites, and the in-situ star
formation is regulated by different physical mechanisms, such as the AGN feedback for massive galaxies and more
stochastic and inefficient star formation for low mass galaxies as discussed above. Thus the two populations of
galaxies, centrals and satellites, are expected to follow very different stellar mass and host halo mass relations.

It would be interesting to investigate the total accreted stellar mass, i.e., those locked in surviving satellites
and those accreted stars which have already been stripped from their parent satellites and are currently in the outer 
stellar halo. Observationally, the calculation of the latter is often achieved through multi-component decomposition 
of galaxy images or surface brightness profiles \citep[e.g.][]{2014MNRAS.443.1433D,2017ApJ...836..115O}. However, 
purely image based two-component decomposition might be dangerous, because the outer component is determined by 
the few outer most data points, which could be sensitive to systematic uncertainties. We postpone more detailed 
investigations to future studies, in which we plan to at first validate the decomposition by conducting 
multi-component fitting to synthetic images based on numerical simulations.

Interestingly, we found indications that blue ICGs tend to have slightly more stellar mass in satellites and also 
higher fractions of stellar mass in satellites versus total stellar mass ($f_\mathrm{sat}$) at $M_{200}\sim10^{12.7}\msun$. 
If robust, this perhaps indicates the late formation time of blue ICGs. $f_\mathrm{sat}$ increases with the increase 
of $M_{\ast,\mathrm{ICG+diffuse}}$ at $M_{\ast,\mathrm{ICG+diffuse}}>10.5$, which is close to 60\% at $\log_{10}M_\ast
/\sun>11.4$ and drops to $\sim$10\% for MW-mass galaxies. At $M_{\ast,\mathrm{ICG+diffuse}}<10.5$, on the other hand,
$f_\mathrm{sat}$ almost does not change with the decrease in $M_{\ast,\mathrm{ICG+diffuse}}$, again implying the more
stochastic star formation of low-mass galaxies and the stochastic accretion of low-mass satellites by low-mass central 
ICGs. In comparison, the relation between $f_\mathrm{sat}$ and $M_{200}$ is more linear over the whole mass range probed.

Our measurements reveal that central ICGs $+$ their stellar halos dominate within $\sim0.15R_{200,\mathrm{mock}}$, 
and $\sim0.15R_{200,\mathrm{mock}}$ marks the start of a transition radius into the satellite dominated region. The 
fractions of total stellar mass versus total mass are the highest in the galaxy center, which keep dropping with the 
increase in projected distances to the central ICGs up to $\sim0.15R_{200,\mathrm{mock}}$. At $r_p>0.15R_{200,\mathrm{mock}}$, 
the stellar mass versus total mass fractions are all below 1\%, and stay almost as a constant, indicating the 
radial distribution of satellites tend to trace the distribution of the underlying dark matter.

In this study, issues related to source deblending within $r_p\sim0.1R_{200,\mathrm{mock}}$ prevent us from proper 
comparisons between the profiles of ICGs $+$ their stellar halos and that of satellites. We leave more careful 
corrections for improper source deblending and multi-component image decomposition, as have been discussed above, 
to future studies. Besides, we did not explicitly distinguish morphology and color in Paper I and this study, and 
it would be interesting to extend our studies to galaxies such as the red spiral galaxies \citep[e.g.][]{2010ApJ...719.1969B,
2019ApJ...883L..36H}, in order to take a closer look at the morphological evolution and its connection to the
quench of star-forming activities and the connection to the host dark matter halos. We also plan to extend our
studies on satellite galaxies, centrals $+$ stellar halos and their host dark matter halos to intermediate and
high redshifts.

\acknowledgments
The Hyper Suprime-Cam (HSC) collaboration includes the astronomical communities of Japan and Taiwan, and Princeton University. 
The HSC instrumentation and software were developed by the National Astronomical Observatory of Japan (NAOJ), the Kavli Institute 
for the Physics and Mathematics of the Universe (Kavli IPMU), the University of Tokyo, the High Energy Accelerator Research 
Organization (KEK), the Academia Sinica Institute for Astronomy and Astrophysics in Taiwan (ASIAA), and Princeton University. 
Funding was contributed by the FIRST program from the Japanese Cabinet Office, the Ministry of Education, Culture, Sports, Science 
and Technology (MEXT), the Japan Society for the Promotion of Science (JSPS), Japan Science and Technology Agency  (JST), the 
Toray Science Foundation, NAOJ, Kavli IPMU, KEK, ASIAA, and Princeton University.

This paper makes use of software developed for the Large Synoptic Survey Telescope. We thank the LSST Project for making their 
code available as free software at  http://dm.lsst.org

This paper is based on data collected at the Subaru Telescope and retrieved from the HSC data archive system, which is
operated by Subaru Telescope and Astronomy Data Center (ADC) at NAOJ. Data analysis was in part carried out with the cooperation
of Center for Computational Astrophysics (CfCA), NAOJ.

The Pan-STARRS1 Surveys (PS1) and the PS1 public science archive have been made possible through contributions by the Institute
for Astronomy, the University of Hawaii, the Pan-STARRS Project Office, the Max Planck Society and its participating institutes,
the Max Planck Institute for Astronomy, Heidelberg, and the Max Planck Institute for Extraterrestrial Physics, Garching, The Johns
Hopkins University, Durham University, the University of Edinburgh, the Queen’s University Belfast, the Harvard-Smithsonian Center
for Astrophysics, the Las Cumbres Observatory Global Telescope Network Incorporated, the National Central University of Taiwan, the
Space Telescope Science Institute, the National Aeronautics and Space Administration under grant No. NNX08AR22G issued through the
Planetary Science Division of the NASA Science Mission Directorate, the National Science Foundation grant No. AST-1238877, the
University of Maryland, Eotvos Lorand University (ELTE), the Los Alamos National Laboratory, and the Gordon and Betty Moore
Foundation.

This paper involves the usage of the astronomical python package of \textsc{astropy} \citep{2013A&A...558A..33A}, the astronomical
source detection software \textsc{Sextractor} \citep{1996A&AS..117..393B}, the python interface of the \textsc{minuit} function
minimizer \citep{1975CoPhC..10..343J} \textsc{iminuit} and the \textsc{emcee} software \citep{2013PASP..125..306F}.

This work is supported by NSFC (12022307, 11973032, 11890691, 11621303), National Key Basic Research and Development Program of 
China (No.2018YFA0404504) and 111 project No. B20019. WW gratefully acknowledge the support of the MOE Key Lab for Particle Physics,
Astrophysics and Cosmology, Ministry of Education. The computation of this work is partly done on the \textsc{Gravity} supercomputer 
at the Department of Astronomy, Shanghai Jiao Tong University. This work was supported in part by World Premier International Research 
Center Initiative (WPI Initiative), MEXT, Japan, and JSPS KAKENHI Grant Numbers JP18H04350, JP18H04358, JP19H00677, JP20H05850, and 
JP20H05855. WW is very grateful for useful discussions with Caina Hao, Lizhi Xie, Xiaoyang Xia and Jie Wang.

%







\appendix

\section{Source deblending issues}
\label{app:deblending}

\begin{figure*}
	\includegraphics[width=0.8\textwidth]{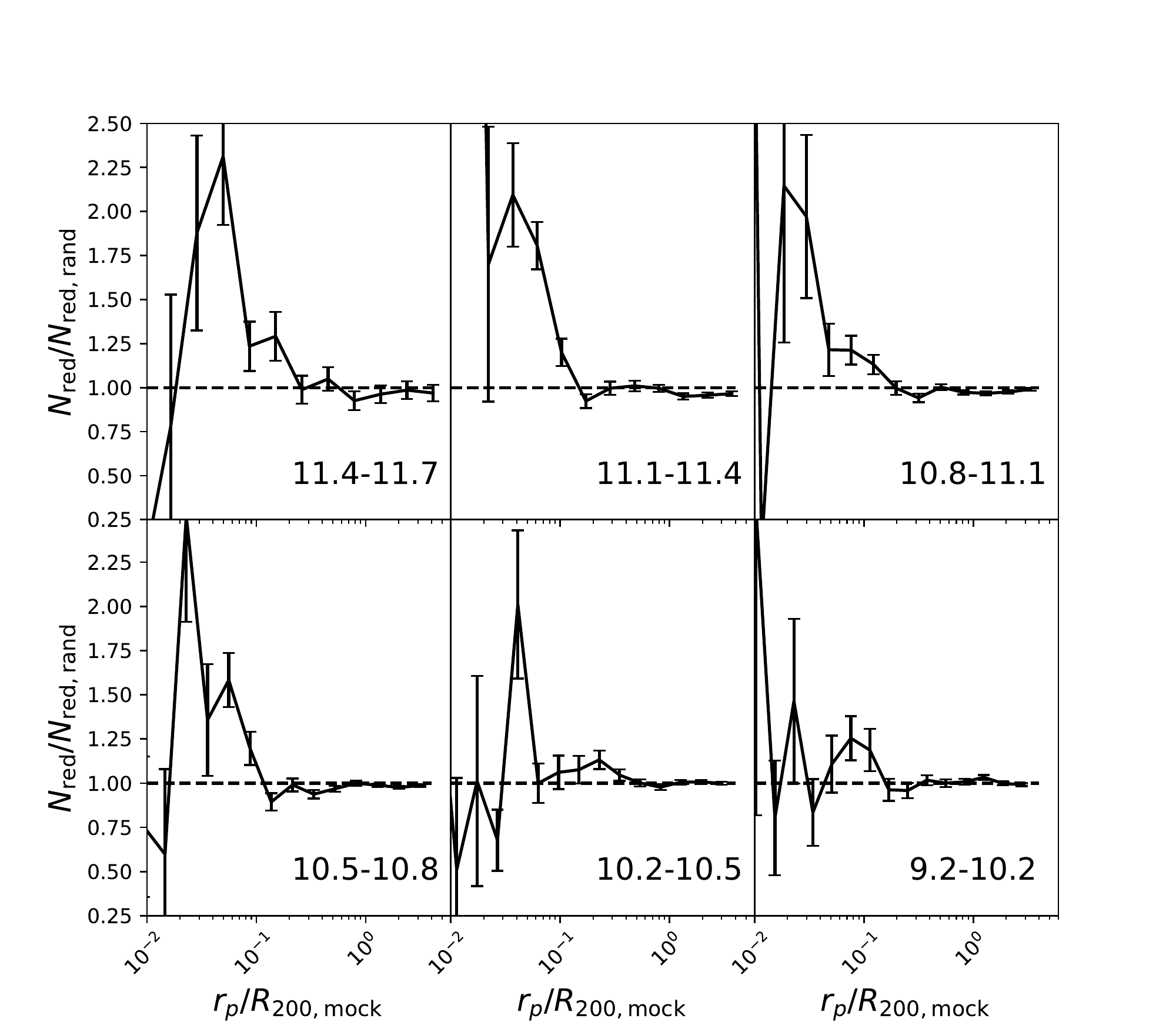}
	\caption{Projected stellar mass density profiles of red background galaxies with $r<25$ around 
	ICGs, divided by the same profiles centered on random points. Measurements on small scales within 
	0.1$R_{200,\mathrm{mock}}$ tend to be significantly affected by source deblending issues.
	}
	\label{fig:profmassbkgd}
\end{figure*}

On scales close to the central ICGs, companion sources might blend with the footprint of the central
galaxy and have to be deblended. However, the radial distribution of satellites may still suffer from
imperfect deblending on scales close to the central dominating ICG. This not only affects satellite 
counts on small scales, but also the surface brightness profiles of ICGs and their stellar halos could 
be affected if companion sources are not properly detected and thoroughly masked (see Section~\ref{sec:SB}). 
For the latter, the problem is less significant because ICGs dominate over smaller companion galaxies. 
The amplitudes and slopes of the projected density profiles for satellites, however, may be significantly 
affected on small scales.

\cite{2012ApJ...751L...5T} corrected their satellite counts on small scales, by modelling the smooth
light distribution of their sample of luminous red galaxies, subtracting it from the image and detecting
remaining sources again. Throughout this paper, we choose to ignore this issue, by focusing on the
radial range which is not affected by deblending issues. This is mainly because of the difficulties
of modelling the light distribution of late-type galaxies, which have rich star-forming regions and
substructures. The choice, however, prevents us from detailed comparison between the radial distribution
of satellites, centrals and the diffuse stellar halos on scales affected by deblending issues for now.
We postpone more detailed corrections for deblending issues to future studies.

To pick up a reasonable inner radius, outside which the satellite counts are not affected by deblending
issues, we count background companions around our ICGs, by using those very red companions with $^{0.1}
(g-r)>0.065\log_{10}M_\ast/\msun+0.35$. Based on their color, if they are real galaxies instead of 
faked sources due to deblending mistakes, it is very unlikely that they stay at the same redshift of the 
central ICGs. These background counts are divided by the counts around random points, and after the division 
the ratios are expected to scatter around unity. Any systematic deviation from unity on small scales mainly 
reflects issues associated with imperfect source deblending. Note source blending happens for fore/background 
sources as well, which does not depend on whether the companion is a true satellite or a background source.

The results are shown in Figure~\ref{fig:profmassbkgd}. To ensure enough number of sources on small scales, we
use all red background sources down to $r=25$. On large scales, the profiles are consistent with unity, whereas 
on small scales close to the central parts of ICGs, the profiles deviate from unity, revealing the deblending 
issues.

In addition to the negative values in the very center, which is probably due to the lacking in sources
within such small projected areas, we see significant positive signals at $r_p/R_{200,\mathrm{mock}}<0.1$. 
We have checked that this cannot be explained by lensing magnifications, which contribute much weaker 
signals. The positive signals could be mainly due to deblending mistakes that parts of the central ICG
are mistakenly deblended to be companions. In addition, we cannot rule out the possibility that maybe 
a very small fraction of real but extremely red satellites may also contribute to such positive signals. 
Despite this, our discussions throughout the main test are focused on the radial region where the profiles 
are not significantly affected by deblending issues ($r_p>0.1R_{200,\mathrm{mock}}$). 

\section{PSF correction}
\label{app:psf}

\begin{figure*}
	\includegraphics[width=0.49\textwidth]{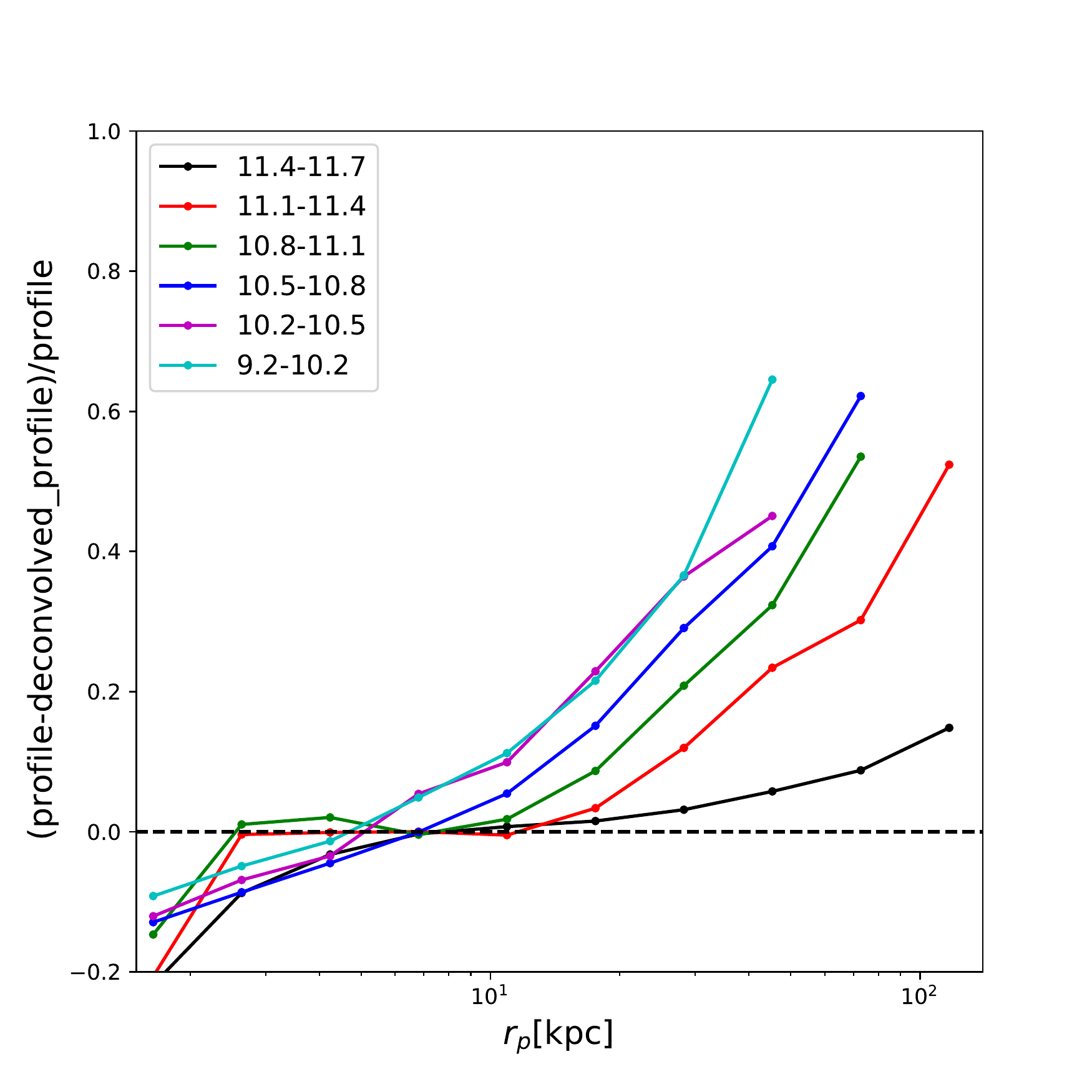}%
	\includegraphics[width=0.49\textwidth]{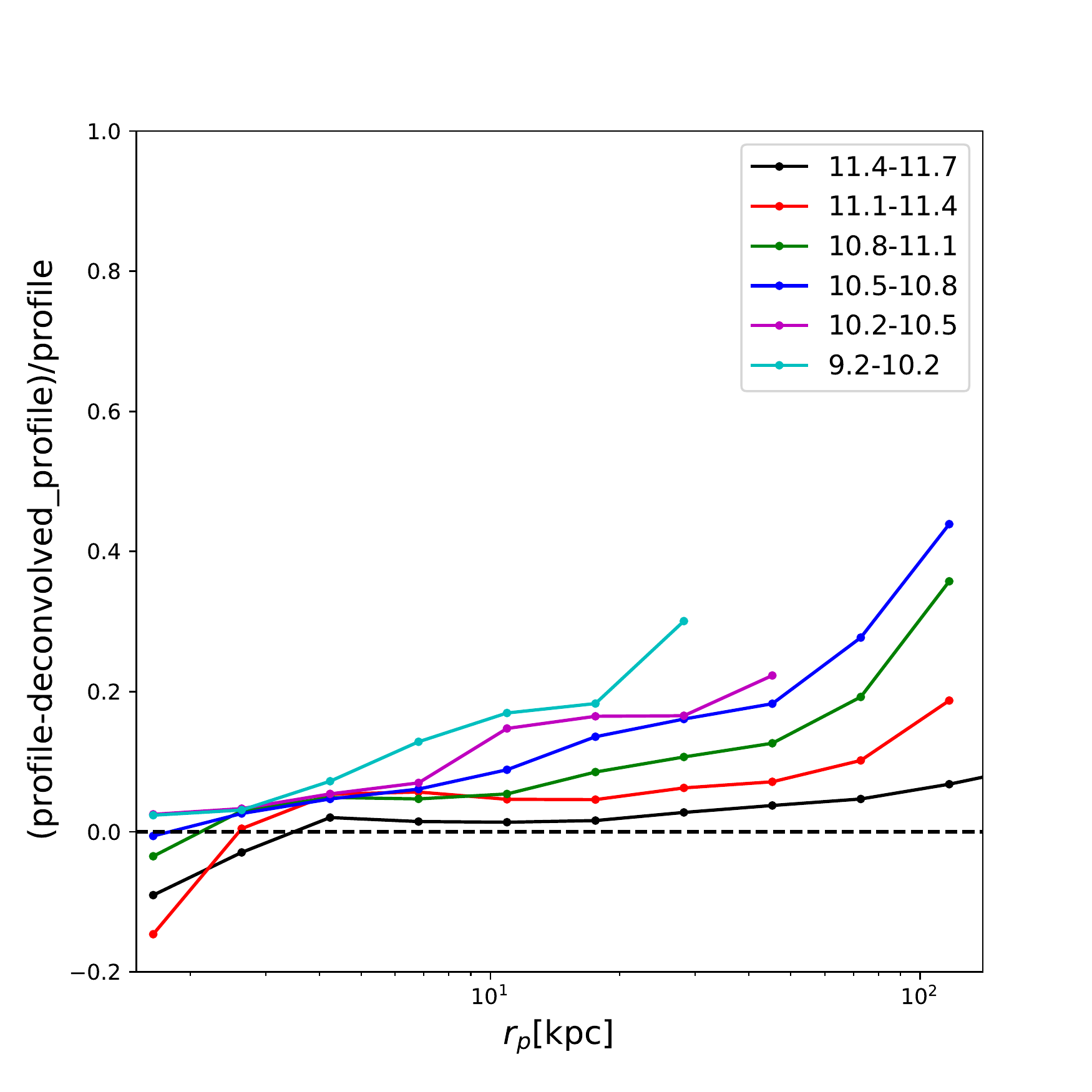}
	\caption{The fractions in the measured surface brightness profiles which are contaminated by PSF
	scattered light for blue (left) and red (right) ICGs in HSC $r$-band, reported as functions of
	the projected radius to the galaxy center, and for ICGs in different log stellar mass bins (see
	the legend).
	}
	\label{fig:psfredblue}
\end{figure*}

In paper I, we measured the extended PSF wings out to $\sim100\arcsec$. Direct deconvolution of the PSF 
is difficult due to noise, and we obtain the PSF-free surface brightness profiles by fitting PSF-convolved 
triple Sersic model profiles to the measured surface brightness profiles of ICGs and their stellar halos 
\citep{2010ApJ...714L.244S,2011ApJ...731...89T}.

It has been shown by \cite{2010ApJ...714L.244S} that the residual added best-fitting models are not sensitive
to variations in the model parameters and is less model dependent, which can be used as estimates of the
PSF-deconvolved profiles, and thus we call them the PSF-corrected profiles.

Figure~\ref{fig:psfredblue} shows the estimated fractions of PSF contamination in HSC $r$-band, for red and 
blue ICGs. Consistent with Paper I, the fraction becomes more significant at larger radii and around 
smaller galaxies, where the outer stellar halos are significantly fainter. In addition, Figure~\ref{fig:psfredblue} 
shows that the PSF contamination has a strong dependence on galaxy color. Red galaxies tend to be contaminated 
less by PSF scattered light at a fixed radius. 

The fractions of PSF contamination have been measured for HSC $g$, $r$ and $i$-bands, but for brevity the $g$ 
and $i$-bands results are not shown. We then correct for the effect of PSF in all three bands, using the fraction 
of PSF contamination for galaxies with the corresponding stellar mass and color. Note, the fraction of PSF 
contamination estimated above is based on stacked profiles, which is on average valid for all galaxies in the 
same bin, whereas there are scatters at individual galaxy level. However, we believe our averaged profiles 
are statistically correct. We can test in the following way. The PSF correction is at first done using the 
PSF contamination fraction estimated from the stacked profile of all ICGs in a given stellar mass bin (this is 
not directly shown in this paper), and then the correction is achieved for red and blue ICGs in the same bin 
separately, using the corresponding fractions for red and blue (Figure~\ref{fig:psfredblue}). We found the 
PSF-corrected profiles are almost identical.

\section{Validating the recovered stellar mass}
\label{app:stellarmasstest}

\begin{figure}
\includegraphics[width=0.7\textwidth]{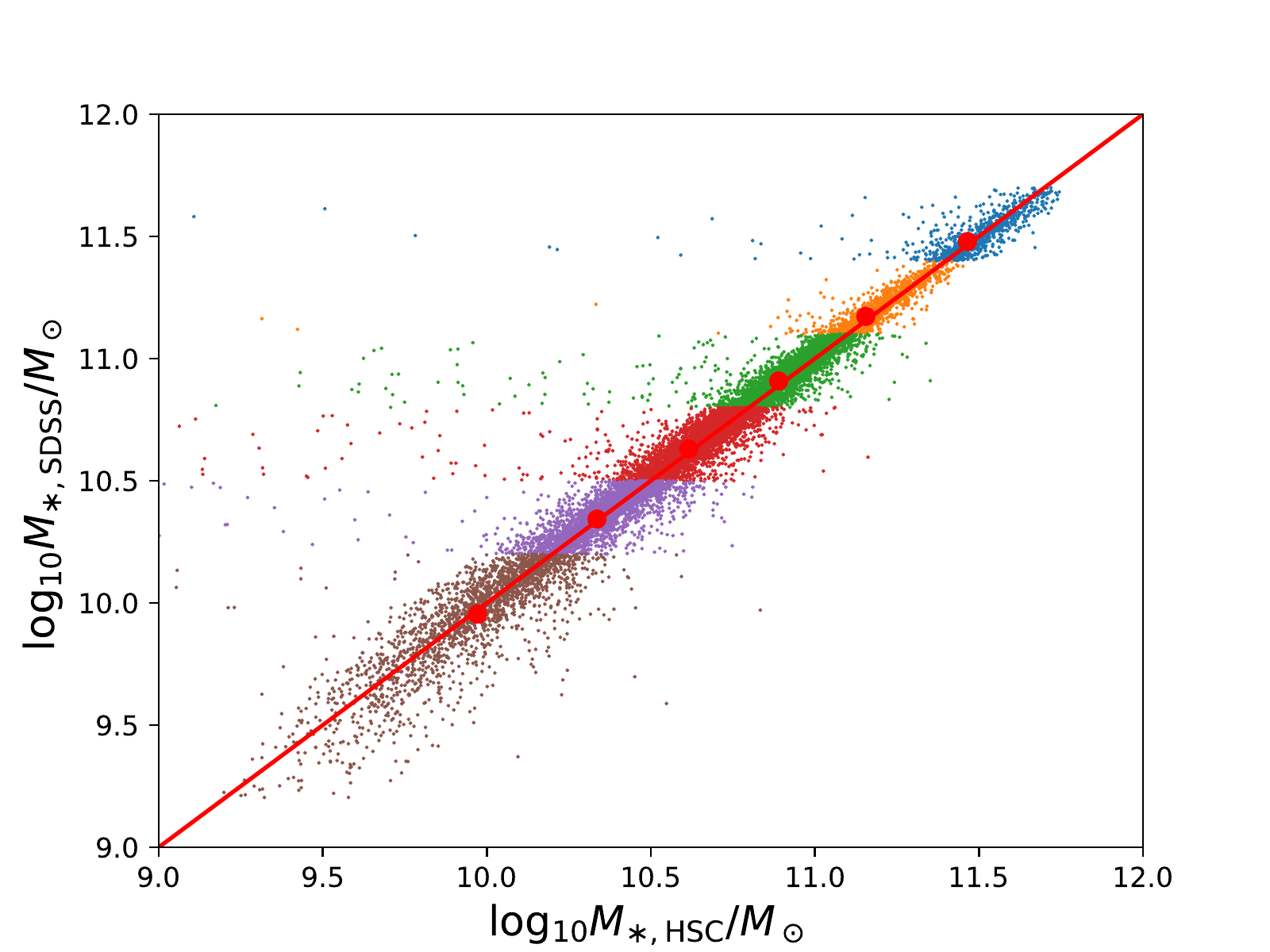}
\caption{The original stellar mass from SDSS ($y$-axis) versus the recovered stellar mass through GPR ($x$-axis), 
for SDSS spectroscopic Main galaxies within the HSC footprint. The recovered stellar mass values are based on the 
$^{0.1}(g-r)$ and $^{0.1}(r-i)$ colors measured from HSC images and are measured within twice the Petrosian radius, 
defined by SDSS.}
\label{fig:scatter}
\end{figure}

\begin{figure}
\includegraphics[width=0.7\textwidth]{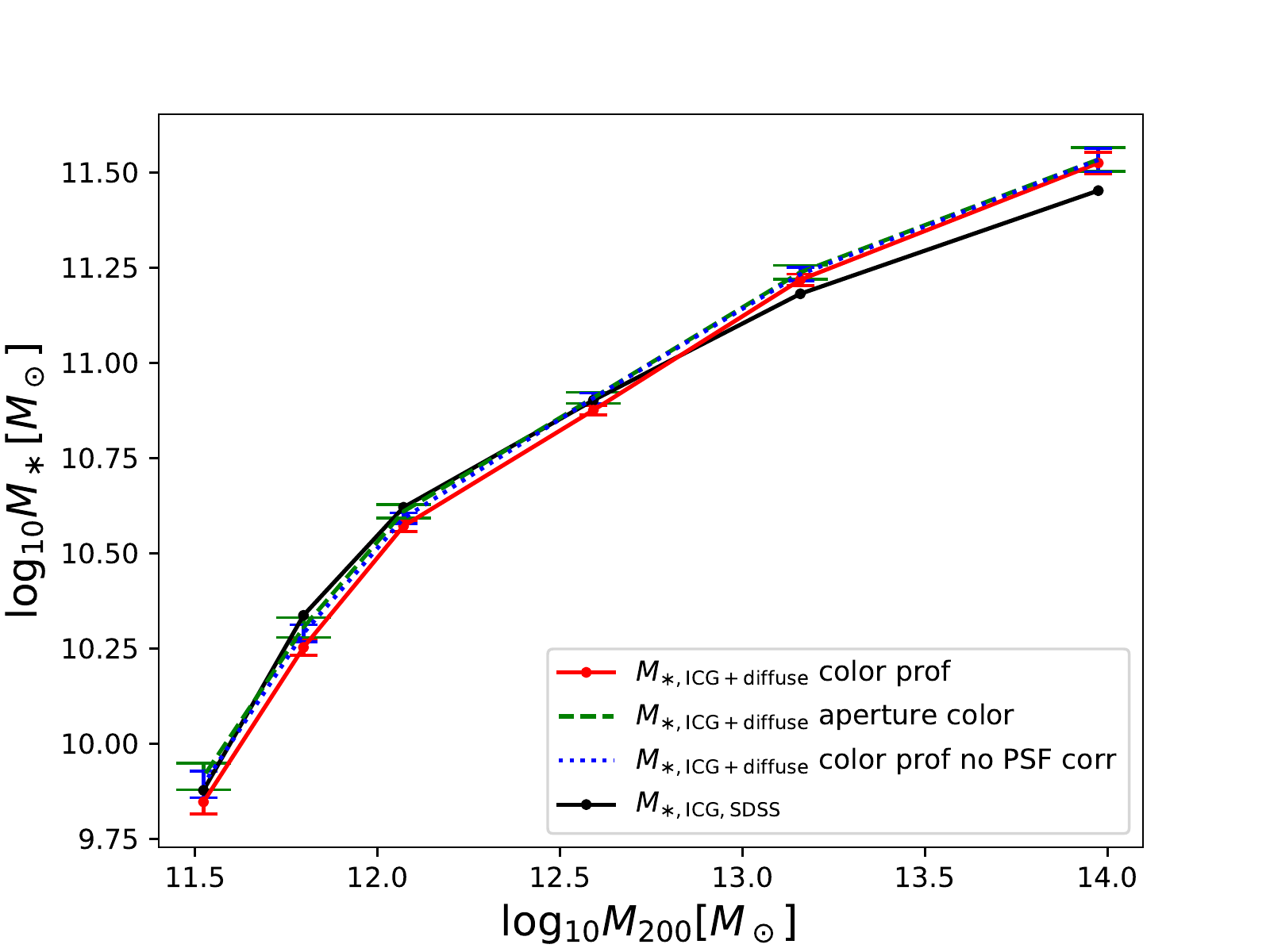}
\caption{Stellar mass defined in a few different ways, reported as functions of the best-fitting $M_{200}$
from weak lensing signals. Black dots connected by lines are the original stellar mass from SDSS (NYU-VAGC). 
Red dots connected by solid lines are the integrated stellar mass over the measured stellar mass density profiles 
of ICGs $+$ their stellar halos, and are based on radius dependent color profiles to infer $M_\ast/L_r$. Green 
dashed lines are similar to the red solid one, but a single aperture color is adopted to infer $M_\ast/L_r$. 
PSF correction has been included for both red and green symbols. Blue dotted lines are similar to the red 
solid one, but without PSF correction. Note for the stellar mass from SDSS, each black dot represents the 
median value for ICGs in the same bin. Mean values lead to slightly different results, but the trends and 
comparisons among different curves remain the same.
}
\label{fig:stellarvshalo}
\end{figure}

For each galaxy image in HSC $gri$-bands, we mask companion sources using a combination of different source
detection thresholds as introduced in Section~\ref{sec:SB}. After masking companions, we calculate the Petrosian
flux of the ICG, which is defined as the total flux within twice the Petrosian radius, and the Petrosian radius
is adopted from SDSS to ensure fair comparisons with SDSS. After applying GPR fitting, the recovered stellar
mass is plotted as the $x$-axis quantity of Figure~\ref{fig:scatter}, while the $y$-axis shows the original
stellar mass from SDSS. Note the SDSS stellar mass is measured by SED fitting to the $ugriz$ Petrosian
flux/magnitude.

It is encouraging to see that the stellar mass recovered from HSC images with GPR agree very well with
the original SDSS stellar mass in the six stellar mass bins (different color), with the mean stellar mass
almost unbiased (red dots). The scatter is flux dependent, ranging from $<0.1$~dex at the bright end to
$\sim0.2$~dex at the faint end. This test not only helps to check the performance of GPR fitting, but
it also reflects the difference between SDSS and HSC photometry within the same aperture. The response
curves of HSC $g$, $r$ and $i$ filters are almost the same as SDSS. However, HSC is significantly deeper
and has smaller PSF size. The way of sky background subtraction and image masks can be quite different
between HSC and SDSS. Despite of these differences, we have validated that we can recover the stellar
mass in an unbiased way.

In Figure~\ref{fig:stellarvshalo}, we show the stellar mass versus halo mass relation, for stellar
mass calculated in a few different ways. The black solid curve corresponds to the original SDSS
(NYU-VAGC) stellar mass. The red solid curve is based on the integrated stellar mass over the projected
stellar mass density profiles in Figure~\ref{fig:profmassall}. The green dashed curve is similar to
the red solid one, but to calculate the projected stellar mass density profile, instead of using
the actual color profiles, we calculate the average color within twice the SDSS Petrosian radius,
and apply the aperture color to the whole radial range to calculate the $M_\ast/L_r$.

Comparing the original SDSS stellar mass with the integrated stellar mass of the red solid curve,
it is clearly shown that at the massive end, the integrated stellar mass is larger than that of SDSS.
At the most massive end, the difference is about 0.07~dex. This is mainly because SDSS is shallower,
and the signals of the outer stellar halo are failed to be detected on individual SDSS images, while
after stacking deep HSC images for galaxies in the same stellar mass bin, we are not only able to go
deeper for the same ICG, but also the background noise level is significantly decreased after stacking,
which enables us to detect signals below the noise level of individual images.

At the low-mass end, the red solid curve is below the black solid one. This is caused by the radial
dependence of $M_\ast/L_r$. The stellar mass of SDSS was obtained through SED fitting to galaxy colors
measured within twice the Petrosian radius. In other words, the same aperture color was applied to
obtain an overall $M_\ast/L_r$ without considering the color gradient. Smaller galaxies have steeper
color gradients, and if adopting a fixed aperture color measured closer to the center of the galaxy,
$M_\ast/L_r$ tends to be over-estimated, which explains the difference between the red solid and
green dashed curves for smaller ICGs, since the central part of galaxy is redder and the $M_\ast/L_r$ 
is larger for red galaxies than blue ones. 

The blue dotted curve is similar to the red solid one, but without including PSF corrections, and 
it seems the integrated stellar mass becomes slightly higher. We have shown in Paper I and Figure~\ref{fig:psfredblue}
that PSF flattens the surface brightness profiles at small radii, while emissions can be scattered 
away from the central ICGs and contaminate the signals of the outer stellar halos, but the integrated 
total flux is expected to conserve with or without PSF. Moreover, Paper I also shows that PSF tends 
to slightly flatten and redden the $g-r$ color profiles in the outer stellar halo, which would result 
in slightly over-estimated $M_\ast/L_r$, and hence may bring in slightly higher values of the integrated 
stellar mass. In fact, the remaining small difference between the green dashed curve and the black solid 
curve for smaller ICGs can be explained by the ignorance of PSF in SDSS.

\section{Validating the empirical $K$-correction}
\label{app:kcorr}

\begin{figure}
\includegraphics[width=0.49\textwidth]{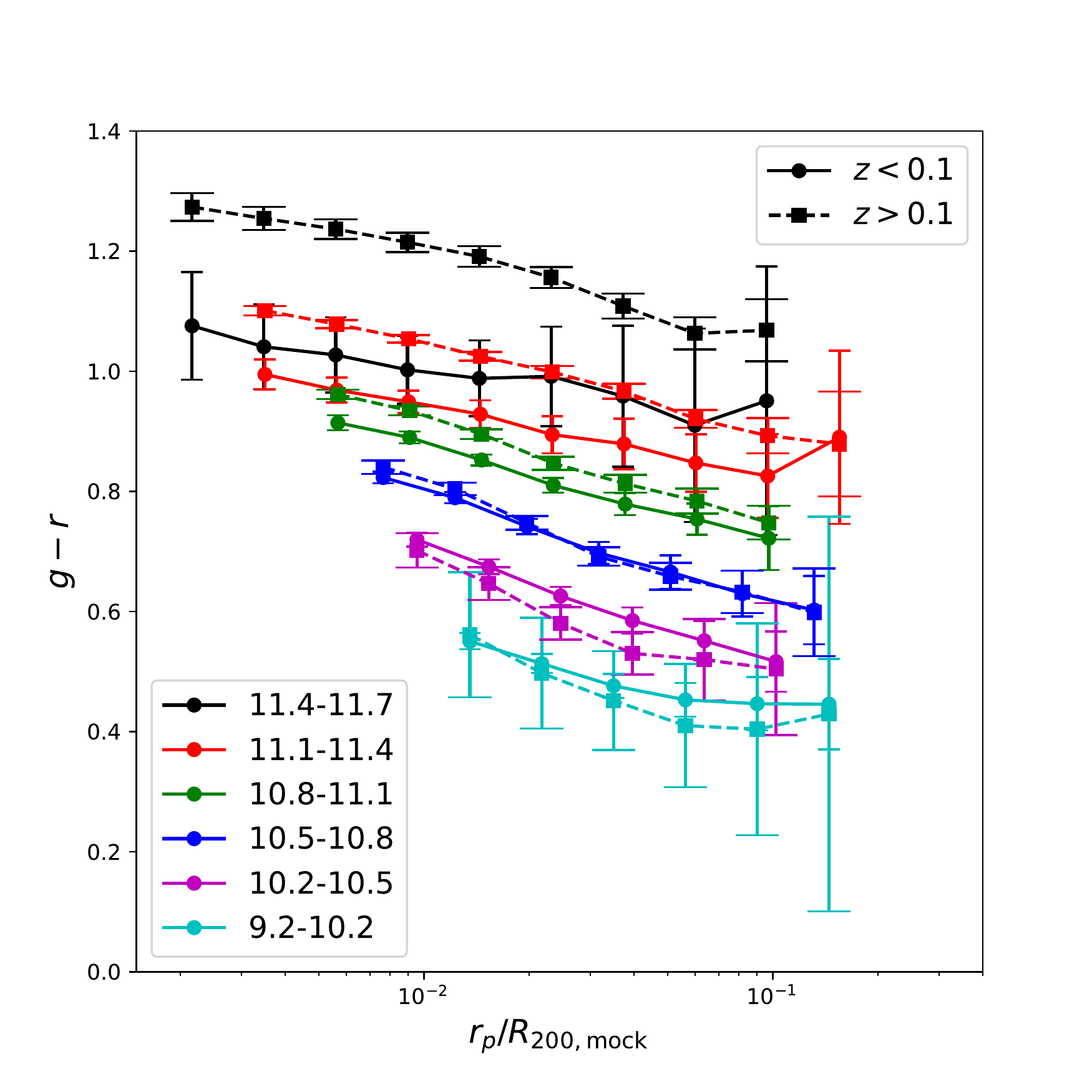}%
\includegraphics[width=0.49\textwidth]{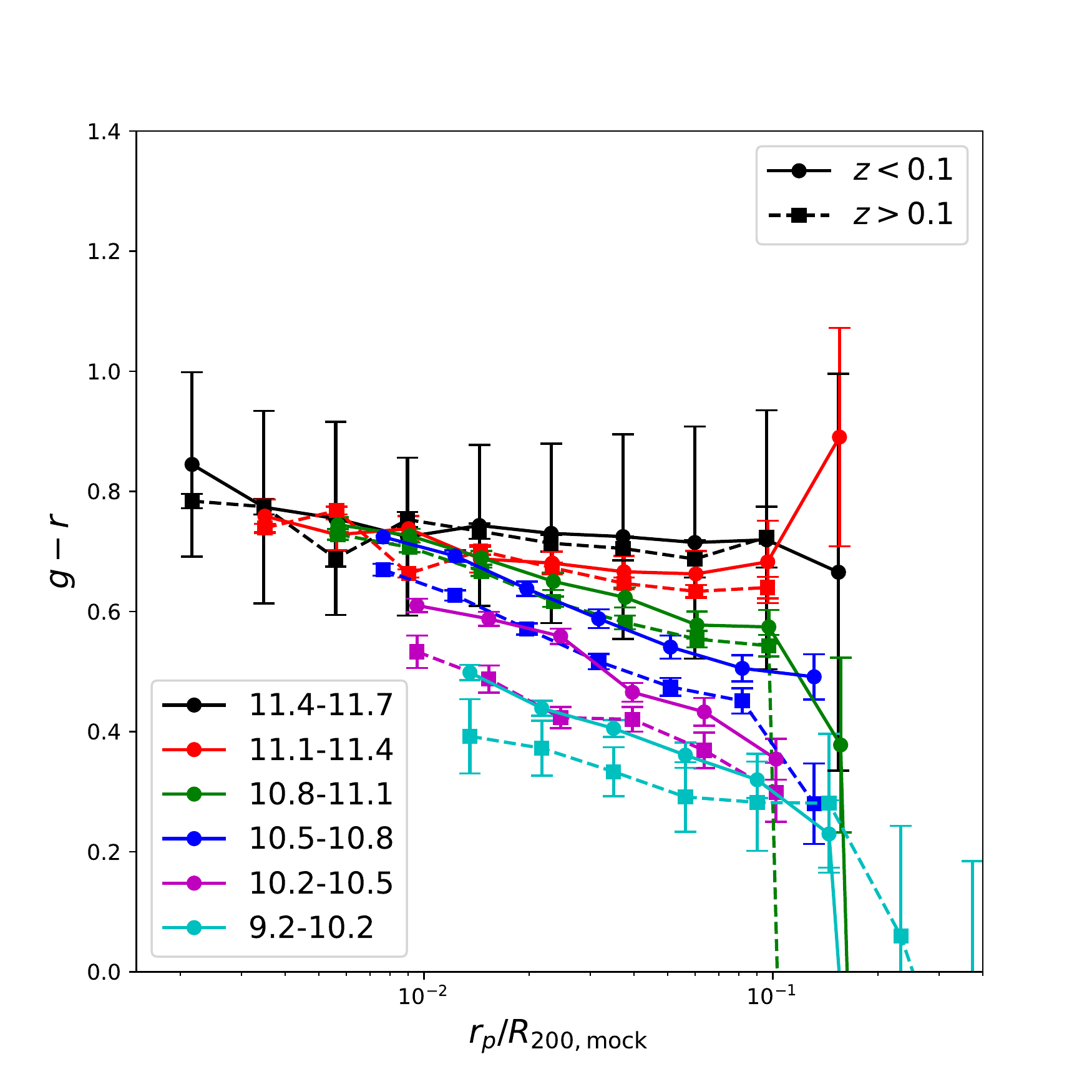}
\caption{$g-r$ color profiles before (left) and after (right) PSF and $K$-corrections, and
for ICGs grouped in a few different stellar mass bins (see the legend).
}
\label{fig:colorevolve}
\end{figure}

In the left plot of Figure~\ref{fig:colorevolve}, we show the color profiles for ICGs in two redshift bins
without including PSF and $K$-corrections. For ICGs more massive than $\log_{10}M_\ast/\msun>10.8$, the higher
redshift subsample tends to have redder colors, which is due to the ignorance of $K$-correction. Massive galaxies
are mostly red, and their passive evolution with redshift is weak, while the prominent 4000{\AA} break feature of
massive galaxies due to absorption shifts to redder bands with the increase in redshift, resulting in prominently
redder colors. For less massive ICGs, they are dominated by star-forming galaxies and have less or no prominent
4000{\AA} break features. We can see the subsample at higher redshifts tends to have bluer colors. This is due to the
selection bias caused by the survey flux limit. At a fixed stellar mass, blue star-forming galaxies have smaller
stellar-mass-to-light ratios, and are brighter. Thus for a fixed flux limit at a given redshift, more small
blue star-forming galaxies can be observed. Hence bluer galaxies tend to be observed in the higher redshift bin.

In the right plot of Figure~\ref{fig:colorevolve}, PSF and $K$-corrections have been included. After $K$-corrections,
massive ICGs at different redshifts now have very similar $g-r$ color profiles ($\log_{10}M_\ast/\msun>11.1$). This
proves that our empirical $K$-correction works reasonably well. Besides, for ICGs less massive than $\log_{10}M_\ast/
\msun\sim11.1$, the higher redshift subsamples all tend to have bluer colors, which are more prominent for smaller
ICGs and also more significant than the left plot. This is because, after eliminating the effect of ignoring
$K$-corrections, the selection bias of a given flux limit, as explained above, becomes more prominent, which
cannot be eliminated after applying $K$-corrections.

\bibliography{master}{}
\bibliographystyle{aasjournal}



\end{document}